\documentstyle[11pt,aaspp4]{article}

 \tighten

\slugcomment{preprint astro-ph/9512121; ApJ, submitted}

\lefthead{Boothroyd \& Sackmann}
\righthead{CNO Isotopes in Red Giants}

     \def\bootplotfidtwo#1#2#3#4#5#6#7#8#9{\centering \leavevmode
       \vbox to#3{\rule{0pt}{#3}}
       \includegraphics{#1}
       \hfil
       \includegraphics{#2}}

\begin{document}

\title{The CNO Isotopes: Deep Circulation in Red Giants
 and First and Second Dredge-up}

\author{Arnold I. Boothroyd\altaffilmark{1}}
\authoremail{aib@krl.caltech.edu}
\affil{Dept.\ of Mathematics \& Statistics, Monash University,
 Clayton, VIC~3168, Australia}
\altaffiltext{1}{now at W.~K.~Kellogg Radiation Laboratory \hbox{106-38},
 California Institute of Technology, Pasadena, CA 91125}

\and

\author{I.-Juliana Sackmann}
\affil{W.~K.~Kellogg Radiation Laboratory \hbox{106-38},
 California Institute of Technology,\\
 Pasadena, CA 91125}
\authoremail{ijs@krl.caltech.edu}

\begin{abstract}

It is demonstrated that {\it deep circulation mixing\/} below the base of
the standard convective envelope, and the consequent
``{\it cool bottom processing\/}'' (CBP) of the CNO isotopes, can reproduce
the trend with stellar mass of the \hbox{$\rm{}^{12}C/{}^{13}C$}
observations in low mass red giants.  (This trend is opposite to what is
expected from standard first dredge-up.)  Our models assume that extra mixing
always reaches to the same distance in temperature from the H-burning shell,
and that CBP begins when the H-burning shell erases the molecular
weight discontinuity (``$\mu$-barrier'') established by first dredge-up.
For Population~I stars, none of the other CNO isotopes except
\hbox{$\rm{}^{15}N$} are expected to be altered by CBP\hbox{}.
(If \hbox{$\rm{}^{18}O$} depletion occurs on the AGB,
as some observations suggest, it would require that extra mixing reach
closer to the H-burning shell on the AGB than on the RGB --- and should also
result in a much lower \hbox{$\rm{}^{12}C/{}^{13}C$} ratio than is observed
in the relevant AGB stars.)

CBP increases dramatically as one {\it reduces\/} the stellar
mass or metallicity --- roughly as $M^{-2}$ on the RGB, due to the longer RGB
of low mass stars, and roughly as~$Z^{-1}$, due to the higher H-shell burning
temperatures of low metallicity stars.
In low mass Population~II stars, {\it all\/} the CNO isotopes
are expected to be significantly altered by CBP\hbox{}.
Field Population~II stars exhibit RGB abundances consistent with the
predictions of our CBP models that have been normalized to
reproduce the Population~I RGB abundances.  On the other hand,
globular cluster stars are observed to encounter much more extensive
processing; additionally, CBP is observed to start near
the base of the globular cluster RGB (overcoming any ``$\mu$-barrier'').

For the CNO isotopes \hbox{$\rm{}^{12}C$}, \hbox{$\rm{}^{13}C$},
\hbox{$\rm{}^{14}N$}, \hbox{$\rm{}^{16}O$}, \hbox{$\rm{}^{17}O$},
and~\hbox{$\rm{}^{18}O$}, we also present self-consistent calculations
of the consequences of both first and second dredge-up, i.e., of standard
convection during the RGB
and AGB stages, over a wide range of stellar masses
(0.8 to $9\>M_\odot$) and metallicities ($Z = 0.02$ to 0.0001).
We demonstrate that the common low and intermediate mass stars are a prime
source of \hbox{$\rm{}^{13}C$}, \hbox{$\rm{}^{14}N$}, and
\hbox{$\rm{}^{17}O$} in the universe.

The light elements (\hbox{$\rm{}^3{He}$}, \hbox{$\rm{}^4{He}$},
\hbox{$\rm{}^7{Li}$}, \hbox{$\rm{}^9{Be}$}, \hbox{$\rm{}^{10}B$},
and~\hbox{$\rm{}^{11}B$}) are discussed in
Sackmann \& Boothroyd (1998a)\markcite{SB98a}.
\end{abstract}



\section{Introduction} \label{intro}

As a star approaches the red giant branch (RGB), its convective envelope
deepens, eventually dredging up products of main sequence nucleosynthesis
({\it first dredge-up\/}).  As the star ascends the RGB, the convective
envelope
first continues to deepen, and then retreats; the point of deepest convection
marks the end of first dredge-up.  Deepest first dredge-up leaves
behind a sharp composition discontinuity.  For low mass stars ($\lesssim
2.5\>M_\odot$), the hydrogen burning shell catches up to and erases this
discontinuity while the star is still on the RGB; for higher masses
($\gtrsim 2.5\>M_\odot$), the star leaves the RGB before this can take
place.  After the completion of core helium burning, the star ascends
the asymptotic giant branch (AGB), and the convective envelope deepens again.
For low mass stars, this is of little consequence, as it does not reach
as deep as first dredge-up; however,
for the higher mass stars ($\gtrsim 4\>M_\odot$
for stars of solar metallicity), it reaches deeper than the
layers mixed by first dredge-up, bringing more nucleosynthesized
material to the surface ({\it second dredge-up\/}).

In the last decade, a wealth of new observations has become available for the
CNO abundances in the envelopes of RGB stars (for Population~I, see, e.g.,
Harris, Lambert, \& Smith 1988\markcite{HarLS88};
Gilroy 1989\markcite{Gil89};
Gilroy \& Brown 1991\markcite{GilB91};
Charbonnel 1994\markcite{Char94};
Charbonnel, Brown, \& Wallerstein 1998\markcite{CharBW98},
and references therein; for Population~II, see, e.g.,
Carbon et al.\ 1982\markcite{Carb+82};
Trefzger et al.\ 1983\markcite{Tref+83};
Langer et al.\ 1986\markcite{Lang+86};
Sneden, Pilachowski, \& VandenBerg 1986\markcite{SnePV86};
Suntzeff \& Smith 1991\markcite{SuntS91};
Kraft et al.\ 1993\markcite{Kra+93}, 1995\markcite{Kra+95},
 1997\markcite{Kra+97};
Pilachowski et al.\ 1993\markcite{PilSB93}, 1997\markcite{Pil+97};
Sneden et al.\ 1994\markcite{Sne+94}, 1997\markcite{Sne+97};
Shetrone 1996a\markcite{Shet96a},b\markcite{Shet96b};
Smith et al.\ 1996\markcite{Smi+96}, 1997\markcite{Smi+97};
and references therein).  Several puzzles arose in interpreting these
observations.  For low mass Population~I stars, many of the observed
\hbox{$\rm{}^{12}C/{}^{13}C$}
ratios were in conflict with the considerably higher values predicted by
standard stellar evolution and nucleosynthesis theory.  For
globular clusters (i.e., low mass Population~II stars), the observed order
of magnitude decline in C/Fe on the RGB was in conflict with theory,
which predicted an almost negligible decline (see, e.g.,
Smith \& Tout 1992\markcite{SmiT92}).
An analogous puzzle has arisen from observations of oxygen isotope ratios
in AGB stars (see, e.g.,
Harris et al.\ 1987\markcite{Har+87},
and references therein), and been accentuated recently by new high-precision
measurements of oxygen isotopic ratios in meteoritic grains which were formed
in circumstellar envelopes, and are believed to have originated in RGB and
AGB stars
(Huss et al.\ 1994\markcite{Hus+94};
Nittler et al.\ 1994\markcite{Nit+94}, 1997\markcite{Nit+97}).
Some of the observed \hbox{$\rm{}^{18}O$} abundances are
much smaller than can be understood in terms of standard
stellar evolution theory.

It has been suggested that the above conflicts might all
be resolved if low mass stars experienced some form of extra deep mixing,
below the conventional convective envelope (see, e.g.,
Dearborn, Eggleton, \& Schramm 1976\markcite{DeaES76};
Sweigart \& Mengel 1979\markcite{SweM79};
Smith \& Tout 1992\markcite{SmiT92};
Charbonnel 1994\markcite{Char94}, 1995\markcite{Char95};
Boothroyd, Sackmann, \& Wasserburg 1995, hereafter BSW95\markcite{BSW95};
Denissenkov \& Weiss 1996\markcite{DenW96}).
This extra mixing would take material from the convective
envelope, transport it down to regions hot enough for some
nuclear processing (in the outer wing of the
H-burning shell), and then transport it back up to the convective envelope;
BSW95\markcite{BSW95}
referred to this as ``{\it cool bottom processing\/}'' (CBP)\hbox{}.
Gilroy \& Brown (1991)\markcite{GilB91},
Charbonnel (1994)\markcite{Char94},
and
Charbonnel et.\ al (1998)\markcite{CharBW98}
demonstrated that the conflict between RGB observations of Population~I
stars and standard theoretical predictions did not arise until after deepest
first dredge-up.  As discussed in \S~\ref{results} below, the situation
appears to be the same for field Population~II stars, although probably not
for globular cluster stars.  Thus one needs a clear understanding of first
dredge-up, before one can understand CBP\hbox{}.

Many first dredge-up calculations have dealt only with solar metallicities,
i.e., Population~I stars (e.g.,
Iben 1965\markcite{Ib65}, 1966a\markcite{Ib66a},b\markcite{Ib66b},
1967\markcite{Ib67};
Dearborn, Tinsley, \& Schramm 1978\markcite{DeaTS78};
Becker \& Cox 1982\markcite{BecC82};
VandenBerg \& Smith 1988\markcite{VandS88};
Landr\'e et al.\ 1990, hereafter La90\markcite{La90};
Dearborn 1992\markcite{Dea92};
Bressan et al.\ 1993\markcite{Bre+93};
El~Eid 1994\markcite{ElE94};
Boothroyd, Sackmann, \& Wasserburg 1994, hereafter BSW94\markcite{BSW94}),
and most investigators have ignored second dredge-up.  There are some
exceptions.  The early work of
Becker \& Iben (1979)\markcite{BecI79}
reported both first and second dredge-up abundances of \hbox{$\rm{}^{12}C$},
\hbox{$\rm{}^{14}N$}, \hbox{$\rm{}^{16}O$}, \hbox{$\rm{}^{18}O$}, and
\hbox{$\rm{}^{22}Ne$} for $Z = 0.001$, 0.01, 0.02, and~0.03, but only for
stellar masses~$\ge 3\>M_\odot$.
Sweigart, Greggio, \& Renzini (1989)\markcite{SweGR89}
reported first dredge-up abundances of \hbox{$\rm{}^{12}C$},
\hbox{$\rm{}^{14}N$}, \hbox{$\rm{}^{16}O$}, \hbox{$\rm{}^{12}C/{}^{13}C$},
C/N, and \hbox{$\rm{}^{16}O/{}^{17}O$} for $Z = 0.004$, 0.01, and~0.04, but
only for masses from~$\sim 1.4$ to~$\sim 2\>M_\odot$ (for masses up
to~$\sim 3\>M_\odot$, they report RGB abundances {\it prior\/} to the
completion of first dredge-up).  The Geneva group
(Schaller et al.\ 1992\markcite{Schal+92};
Schaerer et al.\ 1993a\markcite{Schae+93a},b\markcite{Schae+93b};
Charbonnel et al.\ 1993\markcite{Char+93}, 1996\markcite{Char+96};
Meynet et al.\ 1994\markcite{Mey+94})
reported first dredge-up abundances (and also early-AGB abundances,
{\it prior\/} to the completion of second dredge-up)
of \hbox{$\rm{}^{12}C$}, \hbox{$\rm{}^{13}C$}, \hbox{$\rm{}^{14}N$},
\hbox{$\rm{}^{16}O$}, \hbox{$\rm{}^{17}O$}, \hbox{$\rm{}^{18}O$},
\hbox{$\rm{}^{20}Ne$}, and~\hbox{$\rm{}^{22}Ne$} for
$Z = 0.001$, 0.004, 0.008, 0.02, and~0.04 for masses between 0.9
and~$120\>M_\odot$ --- unfortunately, they provided abundance mass fractions
only to six decimal places, insufficient for obtaining accurate isotope
ratios at low metallicity (at worst, for $Z = 0.001$, \hbox{$\rm{}^{17}O$}
and \hbox{$\rm{}^{18}O$} abundances were typically reported as ``0.000000''
or ``0.000001'').  Note however that the first dredge-up values of
\hbox{$\rm{}^{12}C/{}^{13}C$} and C/N for these models were reported by
Charbonnel (1994)\markcite{Char94}.

Since many observations deal with field Population~II stars and
globular clusters, and with Magellanic Cloud stars
(of intermediate metallicity), we have carried out first and second dredge-up
calculations of the CNO isotopes for a wide range of metallicities.  Note
that the second dredge-up models were standard ones; they do not take into
account any effects that extra mixing might have had in earlier stages of
evolution in low mass stars.

Wasserburg, Boothroyd, \& Sackmann (1995, hereafter WBS95)\markcite{WBS95}
computed parametric models of CBP
in solar-metallicity stars, which suggested that
the \hbox{$\rm{}^{12}C/{}^{13}C$} puzzle on the RGB and the
\hbox{$\rm{}^{18}O$} puzzle on the~AGB might both be resolved.
Charbonnel (1995)\markcite{Char95}
and
Denissenkov \& Weiss (1996)\markcite{DenW96}
used a diffusive mixing algorithm to compute CBP
models of certain isotopes in low-mass Population~II stars; the results
presented in this paper in general agree with theirs.
In the present paper, we use the results of computations similar to those of
WBS95\markcite{WBS95}
to estimate how CBP would alter the first dredge-up
CNO abundances, and present a simple estimate of the relative strength of
CBP as a function of stellar
mass and metallicity.  Similar results for the light elements
\hbox{$\rm{}^3{He}$}, \hbox{$\rm{}^4{He}$}, \hbox{$\rm{}^7{Li}$},
\hbox{$\rm{}^9{Be}$}, \hbox{$\rm{}^{10}B$}, and~\hbox{$\rm{}^{11}B$}
are presented in
Sackmann \& Boothroyd (1998a)\markcite{SB98a}.

For the CNO isotopes,
we have computed estimates of the enrichment of the interstellar
medium from first and second dredge-up and CBP in stars
of near-solar metallicity,
relative to the enrichment resulting from supernovae.


\section{Methods} \label{methods}

We considered stars of 38 different masses from 0.8 to~$9.0\>M_\odot$,
evolving them self-consistently from the pre-main sequence through
first and second dredge-up
until the first helium shell flash (for low and
intermediate mass stars), or to the point where
the program failed due to core carbon
ignition during second dredge-up (for higher masses).
Note that, in low mass stars, we use ``second dredge-up'' to
indicate the deepest penetration of the convective envelope on the
early AGB --- this is a modification of the standard
terminology, in which one would say that second dredge-up did not occur if
the convective envelope reached less deep on the AGB than on the RGB
(as is the case in low mass stars).  For evolutionary program details, see
Boothroyd \& Sackmann (1988)\markcite{BS88},
Sackmann, Boothroyd, \& Fowler (1990)\markcite{SBF90},
and Sackmann, Boothroyd, \& Kraemer (1993)\markcite{SBK93}.
Note that the rezoning algorithm has
been refined to track the composition profiles more
accurately.  This caused only minor changes in our results, the largest
change being in the \hbox{$\rm{}^{12}C/{}^{13}C$} ratio produced by first
dredge-up: in WBS95\markcite{WBS95}, our values for this ratio lay near
the bottom of the range of theoretical calculations by various investigators
(as shown in Fig.~1 of WBS95\markcite{WBS95}), but with the refined rezoning
our values lie near the middle of this ``theoretical range'' (as may be
seen from Fig.~\ref{figcbpc13} below).

Cool bottom processing (CBP) was computed as follows (in a manner similar to
WBS95\markcite{WBS95}).
The envelope structure was taken from a full stellar model, upon
which was superimposed extra mixing (using a ``conveyor-belt'' circulation
model).  We assumed that the extra mixing reached down into the outer wing
of the H-burning shell; the temperature difference $\Delta\log\,T$
between the bottom of mixing and the bottom of the H-burning shell
was considered a free parameter, to be determined by comparison with the
observations.  It was assumed that the value of $\Delta\log\,T$ remained
constant during the evolution up the RGB.  The changes in the envelope
structure were followed as the star ascended the RGB (``evolving RGB''
CBP models, unlike the ``single episode'' CBP models of
WBS95\markcite{WBS95},
where it was assumed that the envelope structure was constant in time).
Envelope structures from full stellar models of mass $1\>M_\odot$ were
used, for $Z = 0.02$, 0.007, 0.001, and~0.0001.  Higher
stellar masses were considered by simply adding more mass to the convective
envelope, and changing to the appropriate first dredge-up abundances and
RGB starting point; this should be a good approximation for the RGB stars
we considered,
namely, those with masses small enough to have degenerate helium cores
(and thus a long RGB).  We assumed that
CBP began at the point on the RGB where the hydrogen shell reached (and
erased) the the molecular weight discontinuity (``$\mu$-barrier'')
that was created at the point of deepest first dredge-up
(Charbonnel 1994\markcite{Char94}).
As in
WBS95\markcite{WBS95},
the free parameter $\Delta\log\,T$ was determined by matching the observed
\hbox{$\rm{}^{12}C$}/\hbox{$\rm{}^{13}C$} ratios for the six
post-RGB stars in~M67, which has solar metallicity, and a main
sequence turn-off mass of $\approx 1.2\>M_\odot$ (in general we ignored
the difference between the turn-off mass and the initial mass of RGB
stars in a cluster --- in M67, where it is largest, this difference should
still be $\le 0.1\>M_\odot$).  The observed
\hbox{$\rm{}^{12}C$}/\hbox{$\rm{}^{13}C$} ratios in M67 range from 11 to~16
(Gilroy 1989\markcite{Gil89};
Gilroy \& Brown 1991\markcite{GilB91}),
comparable to observational error; if this scatter were real, it
would correspond to a range $0.252 \le \Delta\log\,T \le 0.272$ in our
``evolving RGB'' CBP models.  The mean observed
ratio of~13 corresponds to a value of $\Delta\log\,T = 0.262$, which was
the value we used in the CBP calculations reported in~\S~\ref{results}.
Note that this is shallower extra mixing than was used in our
earlier ``single episode'' models of CBP on the RGB
(WBS95\markcite{WBS95}),
which used $\Delta\log\,T = 0.17$; these ``single episode'' CBP models
yield the observed \hbox{$\rm{}^{12}C$}/\hbox{$\rm{}^{13}C$} ratios
after a mixing episode lasting $\sim 1.25 \times 10^7\/$yr on the lower
RGB, while the ``evolving RGB'' CBP models of the present work do not
attain the observed ratios until the tip of the RGB\hbox{}.
The tight constraint on $\Delta\log\,T$ in our models is due to the
strong temperature dependence of the CNO burning rates --- a change
in $\Delta\log\,T$ of only~0.01 results in a change
of $\sim 40$\% in the amount of CNO processing.  For the
evolutionary stages considered here, with thin H-burning shells,
parameterizing the
effective depth of extra mixing by a constant $\Delta\log\,T$ value
is roughly equivalent to assuming that extra mixing always reaches
down to the point with the same molecular weight gradient (see also
Charbonnel et.\ al 1998\markcite{CharBW98}).

We used the OPAL 1995 interior opacities
(Iglesias \& Rogers 1996\markcite{IglR96}),
with Alexander molecular opacities
(Alexander \& Ferguson 1994\markcite{AlexF94};
Tout 1997\markcite{Tout97})
at low temperatures; these latter require a value of $\alpha = 1.67$
(where $\alpha$ is the ratio of the convective mixing length to the
pressure scale height) to obtain a correct solar model
(Sackmann et al.\ 1990\markcite{SBF90}, 1993\markcite{SBK93}).
Tests were also made using older opacity tables.
Interior opacities from the Los Alamos
Opacity Library (LAOL: from Keady [1985]\markcite{Kea85})
yielded only slightly different amounts of dredge-up (see~\S~\ref{results}),
while molecular opacities from Sharp (1992)\markcite{Shar92}
or from Keady (1985)\markcite{Kea85}
required a value of $\alpha = 2.1$ but had no effect on dredge-up.
(Varying $\alpha$ alone has almost no effect on the depth of
dredge-up, as has already been noted by Charbonnel [1994]\markcite{Char94}).


\begin{table*}[!b]

\caption{Initial Isotopic Compositions of Our Models, as a Function of
 Metallicity~$Z$}

 \label{tblinitxi}

\begin{center}
\tabcolsep=0.3 em
\begin{tabular*}{\hsize}{@{}@{\extracolsep{\fill}}*{10}{c}@{}}
\tableline\tableline
     \noalign{\smallskip}
 & & interior & \multicolumn{4}{c}{\dotfill mass fractions\dotfill} &
  \multicolumn{3}{c}{\dotfill number ratios\dotfill} \\
 Z & [Fe/H] & opacities & $Y$ & $C/Z$ & $N/Z$ & $O/Z$ &
  $\!$\hbox{$\rm{}^{12}C/{}^{13}C$} &
  \hbox{$\rm{}^{16}O/{}^{17}O$}$\!$ & \hbox{$\rm{}^{16}O/{}^{18}O$} \\
     \noalign{\smallskip}
\tableline
     \noalign{\smallskip}
   0.02\tablenotemark{a} &
            $0.0$ & {\bf OPAL}, LAOL & 0.280 & 0.1733 & 0.0532 & 0.4823 &
     90 &    2660 &    500 \\
        &   $0.0$ & LAOL             & 0.280 & 0.2179 & 0.0531 & 0.4816 &
     90 &    2660 &    500 \\
\noalign{\medskip}
  0.012 & $-0.35$ & {\bf OPAL}       & 0.264 & 0.1329 & 0.0408 & 0.5617 &
    180 &    5307 &    998 \\
        & $-0.32$ & LAOL             & 0.264 & 0.1817 & 0.0443 & 0.5675 &
    180 &    5307 &    998 \\
\noalign{\medskip}
  0.007\tablenotemark{a} &
           $-0.7$ & {\bf OPAL}       & 0.254 & 0.0952 & 0.0292 & 0.6360 &
    402 &   11885 &   2233 \\
        &  $-0.7$ & LAOL             & 0.254 & 0.1418 & 0.0345 & 0.6625 &
    402 &   11885 &   2233 \\
\noalign{\medskip}
  0.003 &  $-1.2$ & {\bf OPAL}       & 0.245 & 0.0764 & 0.0234 & 0.6728 &
   1225 &   44360 &   6806 \\
        &  $-1.2$ & LAOL             & 0.245 & 0.1067 & 0.0260 & 0.7460 &
   1225 &   44360 &   6806 \\
\noalign{\medskip}
  0.001\tablenotemark{a} &
           $-1.7$ & {\bf OPAL}, LAOL & 0.240 & 0.0764 & 0.0234 & 0.6728 &
   3582 &  106000 &  19910 \\
        &  $-1.6$ & LAOL             & 0.240 & 0.1067 & 0.0260 & 0.7460 &
   3582 &  106000 &  19910 \\
        &  $-1.6$ & LAOL\tablenotemark{b} & 0.240 & 0.1067 & 0.0260 & 0.7460 &
     90 &  106000 &  19910 \\
        &  $-1.3$ & LAOL\tablenotemark{b} & 0.240 & 0.2179 & 0.0531 & 0.4816 &
     90 &    2660 &    500 \\
\noalign{\medskip}
 0.0001\tablenotemark{a}
        &  $-2.7$ & {\bf OPAL}       & 0.238 & 0.0764 & 0.0234 & 0.6728 &
  35820 & 1060000 & 199100 \\
        &  $-2.6$ & LAOL             & 0.238 & 0.1067 & 0.0260 & 0.7460 &
  35820 & 1060000 & 199100 \\
     \noalign{\smallskip}
\tableline
     \noalign{\vskip -30.7pt}
\end{tabular*}
\end{center}

\tablenotetext{a}{\ ``Evolving RGB'' CBP runs were performed for these
metallicities.}

\tablenotetext{b}{\ To give some insight into effects of uncertainties in
initial stellar isotope ratios, these cases test the effect (for $Z = 0.001$)
of assuming that some or all of the initial CNO isotope ratios are
independent of metallicity.}

\end{table*}
\placetable{tblinitxi}

We considered a number of metallicities,
with initial compositions matching those of the corresponding interior
opacity tables in most cases, as shown in Table~\ref{tblinitxi}.  At solar
metallicity ($Z = 0.02$), we used solar system values for the initial carbon
and oxygen isotope ratios and a helium mass fraction~$Y = 0.28$; the ratios
(by mass) $C/Z$, $N/Z$, and~$O/Z$ for the OPAL opacity case were those of
Grevesse \& Noels (1993)\markcite{GreN93}
(for the LAOL opacity case, these abundances were close to
Ross \& Aller [1976]\markcite{RosA76}
or
Grevesse [1984]\markcite{Gre84}).
For lower metallicities,
we reduced the helium abundance via $\Delta Y \approx 2 \Delta Z$
and increased the oxygen content, approximating the observed trend by
$\rm [O/Fe] = -0.5 [Fe/H]$ for $\rm [Fe/H] \ge -1$, and
constant $\rm [O/Fe] = +0.5$ for $\rm [Fe/H] < -1$; observations suggest
that [C/Fe] and [N/Fe] are independent of metallicity (see
Timmes, Woosley, \& Weaver 1995\markcite{TimWW95},
and references therein).
Note that $C/Z$ and $N/Z$ are {\it not\/} constant, since the variation in
[O/Fe] means that $Z$ is not linearly related to the iron
abundance~Fe/H\hbox{}.  Instead, one obtains
\begin{equation}
Z = \cases{
  \{Z_\odot + (10^{0.5}-1) \alpha_\odot\} \,
      10^{\raise1pt\hbox{$\scriptstyle\rm [Fe/H]$}} \> ,
    & ${\rm [Fe/H]} \le -1$ \cr
  (Z_\odot - \alpha_\odot) \, 10^{\raise1pt\hbox{$\scriptstyle\rm [Fe/H]$}} +
      \alpha_\odot \, 10^{\raise1pt\hbox{$\scriptstyle\rm 0.5[Fe/H]$}} \> ,
    & ${\rm [Fe/H]} > -1 \> $, \cr }
\label{eqzfeh}
\end{equation}
where $Z_\odot$ is the solar metallicity, and $\alpha_\odot$ is the mass
fraction in the Sun of those ``$\alpha$-elements''
(\hbox{$\rm{}^{16}O$},~\hbox{$\rm{}^{20}Ne$}, \hbox{$\rm{}^{24}Mg$},~$\ldots$)
that are enhanced relative to Fe
at low metallicity.  For the OPAL cases, opacities and compositions were
interpolated at constant~$Z$ between standard OPAL tables (with
$\rm [O/Fe] = 0$) and an $\alpha$-enhanced OPAL table (``W95hz'', with
$\rm [O/Fe] = 0.5$, and other $\alpha$-enhancements as specified by Weiss:
$\rm [Ne/Fe] = [Si/Fe] = [S/Fe] = 0.3$, $\rm [Mg/Fe] = 0.4$,
$\rm [Ca/Fe] = 0.5$, and $\rm [Ti/Fe] = 0.6$ --- effective
$\alpha_\odot \approx 0.65 \, Z_\odot$).
For the LAOL cases, no $\alpha$-enhanced opacities were available, and
$\alpha_\odot = O_\odot = 0.4816 \, Z_\odot$ was used
(see Table~\ref{tblinitxi}).  The
isotopic ratios \hbox{$\rm{}^{13}C/{}^{12}C$}, \hbox{$\rm{}^{17}O/{}^{16}O$},
and \hbox{$\rm{}^{18}O/{}^{16}O$} were assumed to be proportional to Fe/H
(Timmes et al.\ 1995\markcite{TimWW95};
Timmes 1995\markcite{Tim95}),
i.e., initial \hbox{$\rm{}^{12}C/{}^{13}C$}, \hbox{$\rm{}^{16}O/{}^{17}O$},
and \hbox{$\rm{}^{16}O/{}^{18}O$} were inversely proportional to Fe/H\hbox{}.

On the RGB and AGB, a
Reimers' (1975)\markcite{Rei75}
wind mass loss $\dot M = - \eta (4\times10^{-13}) LR/M$ was included
(where luminosity~$L$,
radius~$R$, and mass~$M$ are in solar units and the mass loss rate~$\dot M$
is in $M_\odot$/yr; $\eta$~is the mass loss parameter).  The value
of~$\eta$ is constrained by globular cluster horizontal branch observations,
which constrain Population~II stars of
initial mass $0.8 - 0.85 \> M_\odot$ to lose between 0.1 and~$0.3 \> M_\odot$
on the RGB, over a metallicity range of 2 orders of magnitude
(Renzini 1981\markcite{Ren81};
Renzini \& Fusi Pecci 1988\markcite{RenF88});
thus $\eta$ must be nearly independent of~$Z$.  We chose
$\eta \propto Z^{0.061}$, yielding
$\eta = \{ 0.435, 0.5, 0.535, 0.563, 0.581, 0.6 \}$
for $Z = \{ 0.0001, 0.001, 0.003, 0.007, 0.012, 0.02 \}$, respectively ---
these $\eta$ values obey the above globular cluster constraint, for both
Alexander and Sharp molecular opacity cases.  In no case had a
significant fraction of
the star's mass been lost by the time first dredge-up was
encountered (or second dredge-up, in intermediate mass stars).  Test runs
with $\eta = 0$, or $\eta = 1.4$ (recommended for intermediate mass
stars by
Kudritzki \& Reimers [1978]\markcite{Kudr78}),
confirmed that the mass loss rate had negligible effect on dredge-up
abundances, except for second dredge-up in low mass stars (where CBP on the
RGB will in any case have altered the composition in a way that our
standard second dredge-up models do not take into account).

Nuclear reaction rates from
Caughlan~\& Fowler (1988; hereafter CF88)\markcite{CF88}
were used, except for the the ${}^{12}{\rm C}(\alpha,\gamma)$
rate, where the rate from
CF88\markcite{CF88}
was multiplied by a factor of~1.7, as recommended by
Weaver \& Woosley (1993)\markcite{WeaW93},
and for the \hbox{$\rm{}^{17}O$}-destruction reactions
${}^{17}{\rm O}(p,\alpha){}^{14}{\rm N}$
and ${}^{17}{\rm O}(p,\gamma){}^{18}{\rm F}
(e^{\hbox{$\scriptscriptstyle +$}}\nu){}^{18}{\rm O}$, where the rates from
Blackmon (1996\markcite{Bl96}, hereafter Bl96)
were used.  The
Bl96\markcite{Bl96}
rates are based on the \hbox{$\rm{}^{17}O$}$(p,\alpha)$ cross section
measurement of
Blackmon et al.\ (1995)\markcite{Bla+95};
they are somewhat higher than the rates of
La90\markcite{La90},
in that the uncertain factor~$f_1 \approx 0.2$ in the
La90\markcite{La90}
formulae for the $\hbox{$\rm{}^{17}O$}+p$ rates is increased to a
value of $f_1 = 0.31 \pm 0.06$ in the
Bl96\markcite{Bl96}
rates.  Note that \hbox{$\rm{}^{17}O$} abundances calculated using the
Bl96\markcite{Bl96}
rates are almost identical to those calculated using the
La90\markcite{La90}
rates (identical below $2\>M_\odot$ and differing by less than 20\% at
higher masses, despite the factor of~1.5 difference in the rates in the
temperature interval $7.4 \lesssim \log\,T \lesssim 8.0$);
both rates thus yield a reasonable fit to the somewhat sparse oxygen isotope
observations, as discussed by
BSW94\markcite{BSW94}
(and references therein).  We also tested the effects on second dredge-up
of using the ${}^{12}{\rm C}(\alpha,\gamma)$ rate from
CF88\markcite{CF88}
(instead of multiplying it by~1.7), or of using the rate from
Caughlan et al.\ (1985)\markcite{CFHZ85}
(which is nearly 3 times larger than that of
CF88\markcite{CF88},
i.e., about another 1.7~times larger than the rate we generally used);
these changes turned out to have little effect.


\section{Results} \label{results}


\subsection{Composition Profiles and Depth of Dredge-up}
 \label{profdepth}


\begin{figure}[!t]
  \plottwo{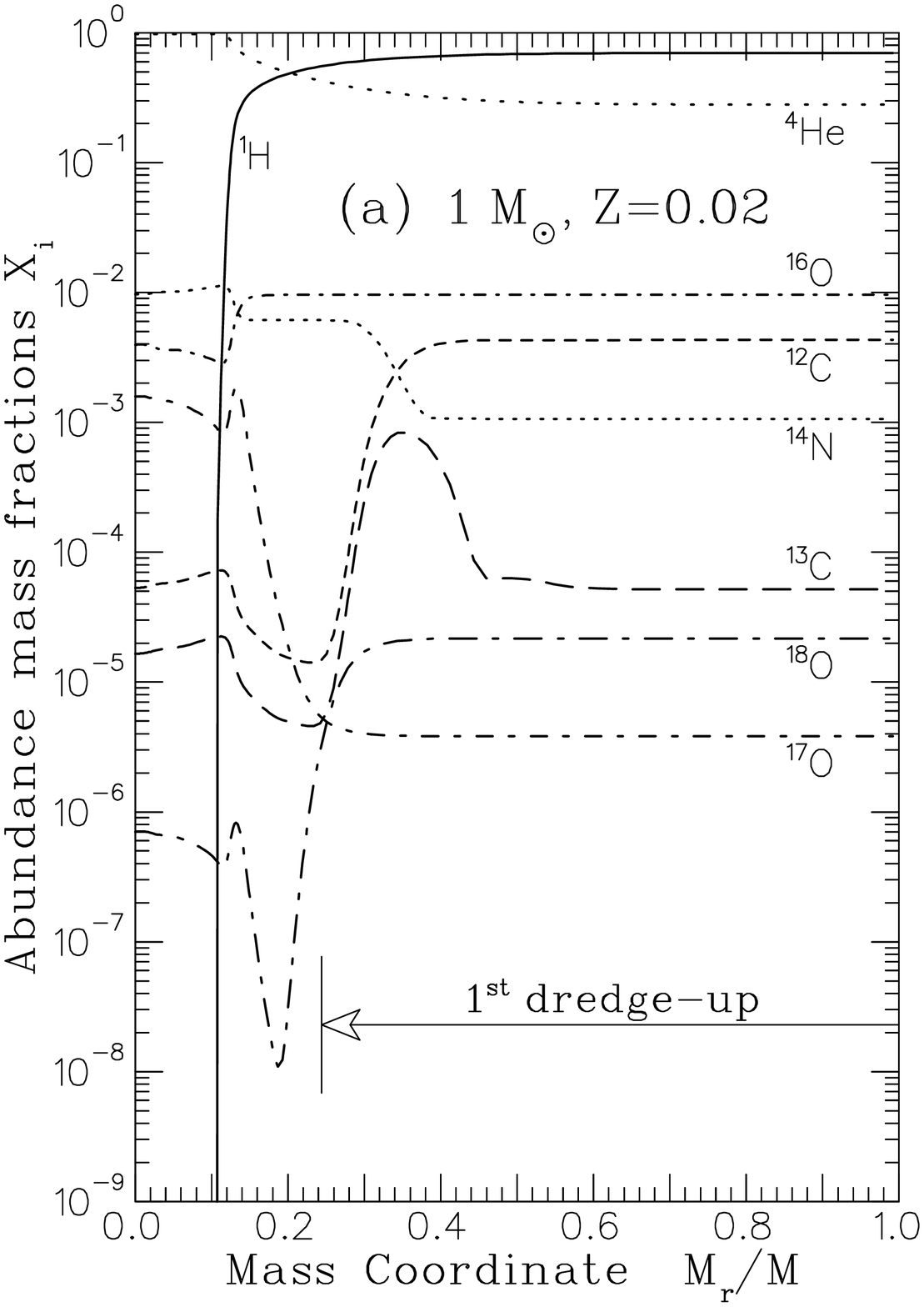}{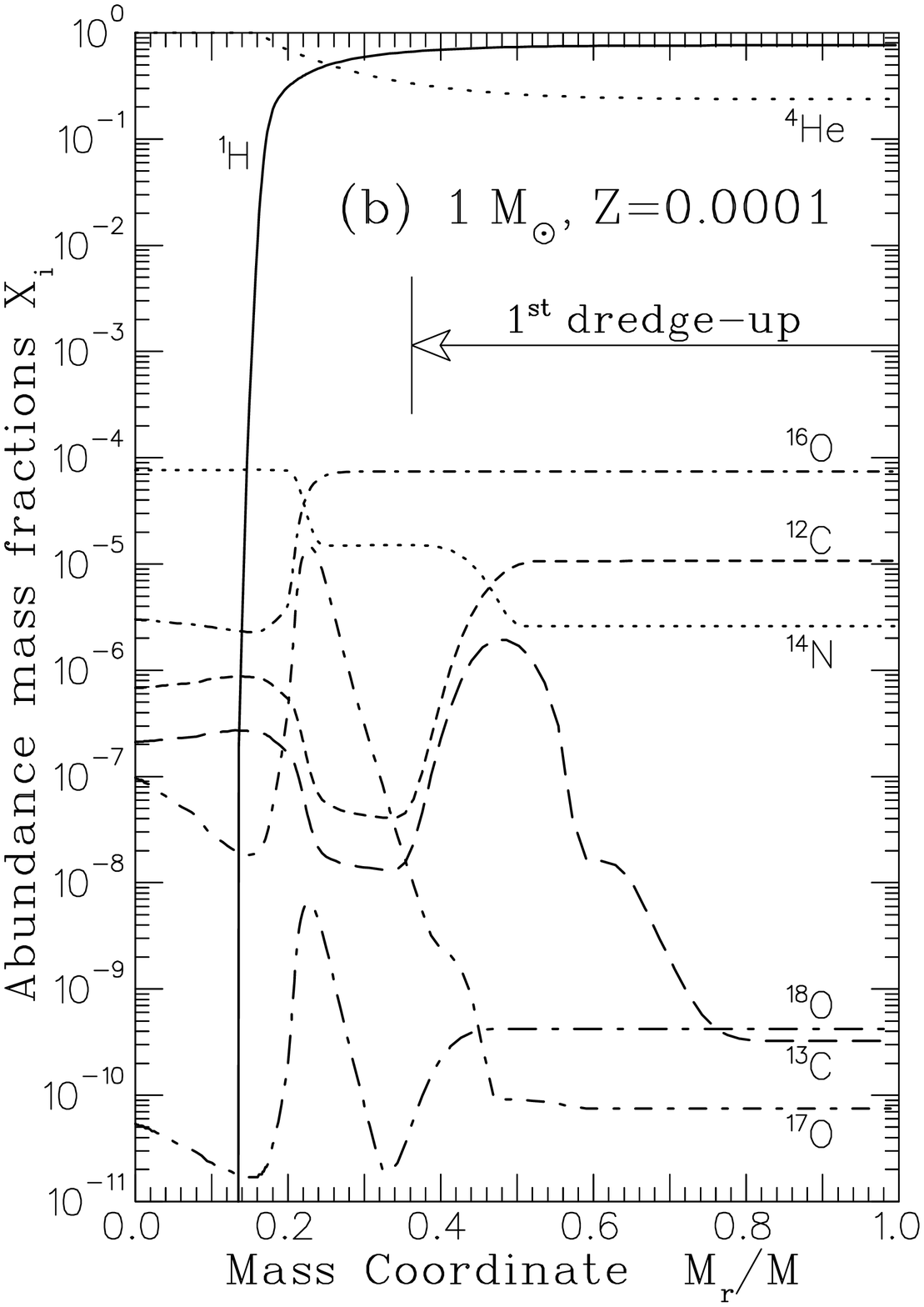}

\caption{Composition profiles as a function of the normalized mass
coordinate~$M_r/M$, for $1\>M_\odot$ stars near the base of the RGB,
prior to first dredge-up; the depth of first dredge-up is indicated by the
horizontal arrow.
(a)~Population~I ($Z = 0.02$);
(b)~Population~II ($Z = 0.0001$).
\label{figdr1}
}

\end{figure}
\placefigure{figdr1}

Figure~\ref{figdr1} presents the composition profiles of $1\>M_\odot$
Population~I and~II stars near the base of the RGB, shortly before first
dredge-up.  Abundance profiles in stars of higher mass are qualitatively
similar, though the peaks and dips in the profiles tend to lie at somewhat
larger values of the normalized mass coordinate~$M_r/M$ (a shift similar
to that from Fig.~\ref{figdr1}a to~\ref{figdr1}b).
The abundance changes produced by first dredge-up (presented
in~\S~\ref{dredge}) can be understood in terms of these profiles.
For example, the relative width of the \hbox{$\rm{}^{13}C$}~pocket increases
somewhat with increased stellar mass in low mass stars, but is almost
constant in intermediate mass stars.  This
explains why first dredge-up \hbox{$\rm{}^{13}C$} enrichment
increases as a function of stellar mass for low masses, and then
levels off for intermediate masses
(the increasing depth of dredge-up with stellar mass is irrelevant, as
the entire \hbox{$\rm{}^{13}C$} pocket is dredged up.)  As pointed out by
Charbonnel (1994)\markcite{Char94},
for Population~I stars the
observed \hbox{$\rm{}^{12}C/{}^{13}C$} ratios in the luminosity range
expected for first dredge-up are in agreement with
the theoretical models of first dredge-up
(see also \S~\ref{dredge}); the subsequent observed
reduction in \hbox{$\rm{}^{12}C/{}^{13}C$} in low mass stars
is not due to further dredge-up, but to cool bottom processing (CBP)\hbox{}.


\begin{figure}[!t]
  \plotfiddle{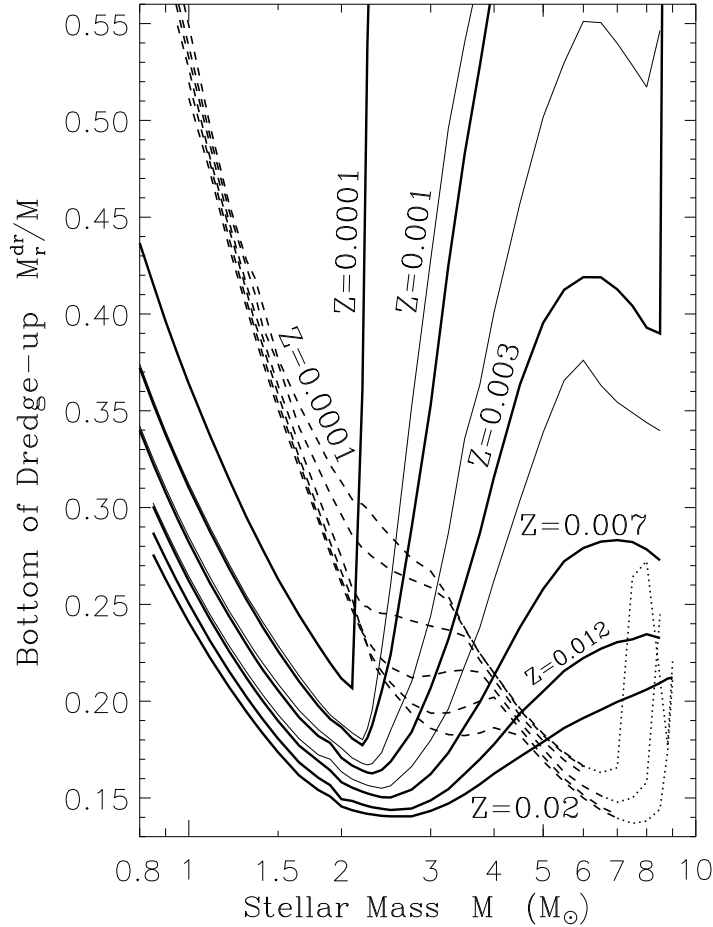}{4.5 true in}{0}{54}{54}{-180}{-43}

\caption{Innermost mass layer~$M_r^{\rm dr}$ reached by the convective
envelope during first dredge-up ({\it solid curves\/}) and during second
dredge-up ({\it dashed curves\/}), as a function of stellar mass~$M$,
for the metallicities of Table~\protect\ref{tblinitxi}
(see~\S~\protect\ref{methods}); for $0.001 \le Z \le 0.007$,
{\it thin solid curves\/} show the effect of using the older LAOL
interior opacities (above $2.5\>M_\odot$, the omitted LAOL $Z = 0.02$
and~0.012 curves would
roughly coincide with OPAL $Z = 0.012$ and~0.007 curves, respectively).
The metallicities are indicated on the solid (first dredge-up)
curves; the metallicities of the dashed (second dredge-up) curves
are in the same order. For $Z = 0.02$, 0.007, and~0.0001, the {\it dotted\/}
continuations of the dashed curves show the depth reached by second
dredge-up during core carbon ignition, in stars massive enough for
carbon burning; second dredge-up may reach deeper subsequently.
\label{figmrdr}
}

\end{figure}
\placefigure{figmrdr}

Figure~\ref{figmrdr} shows the depth in mass of first and second dredge-up,
over a wide range of metallicities.
For solar metallicity, our depths of first dredge-up agree with those
computed by
El~Eid (1994)\markcite{ElE94}
and
Charbonnel (1994)\markcite{Char94};
for $Z = 0.001$, our $1.25\>M_\odot$ case agrees with Charbonnel's, but
we find significantly shallower first dredge-up for~$5\>M_\odot$
($M_r^{\rm dr}/M = 0.7$ for OPAL opacities or 0.75 for LAOL, as opposed
to her value of~0.425, where $M_r^{\rm dr}$ is the deepest mass layer
in the star reached by the convective envelope during dredge-up).  This
indicates that the depth of first dredge-up in
low-metallicity intermediate mass stars is sensitive to the physical inputs
of the stellar models (e.g., equation of state,
opacities, and nuclear rates) that determine the
onset of core helium burning and thus the end of the RGB\hbox{}.
Fortunately, such differences are erased by second dredge-up.
Second dredge-up in low mass stars is significantly shallower
than first dredge-up, and thus would be expected to have
negligible effect on their surface composition (though this {\it may\/} not
always be true: see~\S~\ref{nodredge}); in intermediate mass stars, where it
reaches deeper than first dredge-up, second dredge-up is of key importance.
Figure~\ref{figmrdr} does not show the effect on second dredge-up of using
the older LAOL interior opacities (rather than OPAL), but the relative
shifts are similar to those of first dredge-up, namely, equivalent to a shift
to the next-lower metallicity case for Population~I stars, and little or
no effect in Population~II stars.  (The actual
shift in the depth of second dredge-up from using the LAOL opacities is thus
always small.)  Changing the low-temperature molecular opacities, the
mixing length parameter~$\alpha$, the ${}^{12}{\rm C}(\alpha,\gamma)$ rate,
the CNO fractions relative to~$Z$, or the mass loss rate had negligible
effect on the depth of dredge-up, as noted in~\S~\ref{methods}.

Note that for our most massive stars ($\gtrsim 7\>M_\odot$), central carbon
ignition takes place during second dredge-up.  We report the depth of second
dredge-up at the point during this stage when our program failed
(see dotted lines in Fig.~\ref{figmrdr}); second
dredge-up might reach deeper during subsequent evolution.
The chemical compositions reported in this paper for these more massive stars
may thus underestimate somewhat the effect of second dredge-up.


\subsection{Envelope Isotope Ratios From Dredge-up and CBP}
 \label{dredge}

In this section, we discuss the effects of standard
first and second dredge-up on the CNO isotopic abundances in stellar envelopes.
We also present the results of our ``evolving RGB'' CBP models (CBP on the
RGB only) for the four metallicities where these computations were performed.
These CBP models were parametric studies, with a free
parameter $\Delta\log\,T$ (giving the location, in temperature, of the
bottom of extra mixing, relative to that of the bottom of the H-burning
shell),
fixed by matching
Gilroy's (1989)\markcite{Gil89}
average observed \hbox{$\rm{}^{12}C/{}^{13}C$}
ratio in the open cluster~M67
(as discussed in~\S~\ref{methods}).
The results are summarized in Tables~\ref{tabdrcbp}--\ref{tabdrcbpvlzo}; more
detailed tables are available from the authors\footnote[1]{\ Send e-mail to
{\tt aib}@{\tt krl.caltech.edu} to obtain these tables; also available at
A.~I.~B.'s Web page: {\tt http://www.krl.caltech.edu/${}^{\sim}$aib/}}.
Note that
Lattanzio \& Boothroyd (1997)\markcite{LatB97}
also present figures showing some of the results
discussed here (using the older LAOL interior opacities).

\placetable{tabdrcbp}

\placetable{tabdrcbpz7o}

\placetable{tabdrcbplzo}

\placetable{tabdrcbpvlzo}


\subsubsection{Carbon Isotopes}
 \label{cdredge}


\begin{figure}[!t]
  \plotfiddle{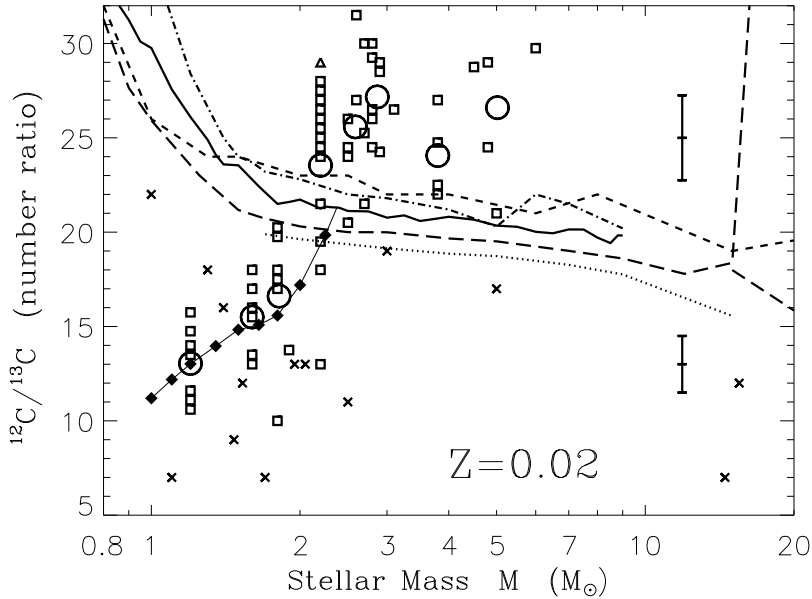}{2.7 true in}{90}{50}{50}{200}{-45}

\caption{Comparison between observations and theory for
\hbox{$\rm{}^{12}C/{}^{13}C$} in solar metallicity red giants.
{\it Open squares\/}: open cluster observations of
Gilroy (1989)\protect\markcite{Gil89},
with {\it open triangle\/} giving lower limit --- {\it large open circles\/}
give mean values at corresponding masses (errorbars at
right show typical uncertainties in individual observations).
{\it Crosses\/}: observations of isolated stars by
Harris \& Lambert (1984a\protect\markcite{HarL84a},b\protect\markcite{HarL84b})
and
Harris et al.\ (1988)\protect\markcite{HarLS88},
where the stellar masses are also uncertain (by a factor of~$\sim 2$).
Theoretical first dredge-up curves:
{\it heavy solid line\/}: present work; {\it dotted\/}:
El~Eid (1994)\protect\markcite{ElE94};
{\it dot-dashed\/}:
Bressan et al.\ (1993)\protect\markcite{Bre+93};
{\it short-dashed\/}:
Dearborn (1992)\protect\markcite{Dea92};
{\it long-dashed\/}:
Schaller et al.\ (1992)\protect\markcite{Schal+92}
(also presented by Charbonnel [1994]\protect\markcite{Char94}),
where second dredge-up is also shown for masses~$\ge 15\>M_\odot$ where
first dredge-up is hard to define.
{\it Filled diamonds\/} (connected by {\it light solid line\/}) indicate
the abundances from the ``evolving~RGB'' CBP models of the present work
(normalized by observations at $1.2\>M_\odot$: see text).
\label{figcbpc13}
}

\end{figure}
\placefigure{figcbpc13}

Figure~\ref{figcbpc13} demonstrates that our ``evolving RGB'' CBP models
can account for the observed trend of \hbox{$\rm{}^{12}C/{}^{13}C$} with
stellar mass.  The field star observations of
Harris \& Lambert (1984a\markcite{HarL84a},b\markcite{HarL84b})
and
Harris et al.\ (1988)\markcite{HarLS88}
are also shown, but their stellar mass values are too uncertain (by a factor
of~$\sim 2$) to be of much use.  There is some scatter in
Gilroy's (1989)\markcite{Gil89}
open cluster \hbox{$\rm{}^{12}C/{}^{13}C$} observations
($\pm 3$ at~$1.2\>M_\odot$); if it
reflects a true scatter in the abundances, this
would imply a variation of order $\pm 40$\% in
the amount of processing resulting from extra mixing at that mass,
corresponding to a depth parameter $\Delta\log\,T$ in the range 0.252 to~0.272
in our ``evolving RGB'' CBP models ($\Delta\log\,T$ is the difference in
temperature between the bottom of extra mixing and the bottom of the
hydrogen shell: see~\S~\ref{methods}).  Note that
Charbonnel et al.\ (1998)\markcite{CharBW98}
obtained a similar depth estimate $\Delta\log\,T \approx 0.26$ (using
\hbox{$\rm{}^{12}C/{}^{13}C$} observations in field Population~I stars
with accurate HIPPARCOS parallaxes); this depth corresponded to a
molecular weight gradient $\nabla \ln\,\mu \sim 1.5 \times 10^{-13}$
in their stellar model, in agreement with the critical $\mu$-gradient
required to explain observed solar lithium and beryllium depletion in
the best solar models of
Richard et al.\ (1996)\markcite{Rich+96}
and
Richard \& Vauclair (1997)\markcite{RichV97}.

Even for intermediate mass stars, where there is no CBP,
there is still some uncertainty in the \hbox{$\rm{}^{12}C/{}^{13}C$} ratio
resulting from first dredge-up.  The observations of
Gilroy (1989)\markcite{Gil89}
for stars of masses $\gtrsim 2.5\>M_\odot$ suggest that the theoretical
models shown in Figure~\ref{figcbpc13} may overestimate the amount
of \hbox{$\rm{}^{13}C$} in the
\hbox{$\rm{}^{13}C$}~pocket by $15 - 30$\%.  Note that an uncertainty in
the depth of dredge-up cannot have any effect, since the entire
\hbox{$\rm{}^{13}C$}~pocket is dredged up, but a smaller
\hbox{$\rm{}^{13}C$}~pocket might possibly result from slight errors in
relative rates of nuclear reactions.  Another possibility is extra rotational
(or diffusional) mixing in the stellar interior on the main sequence,
which is not included in standard stellar models; this type of extra mixing
is usually invoked to explain main sequence \hbox{$\rm{}^7{Li}$} depletion
(see, e.g.,
Vauclair 1988\markcite{Vau88};
Pinsonneault et al.\ 1989\markcite{Pin+89}).


\begin{figure}[!t]
  \plotfiddle{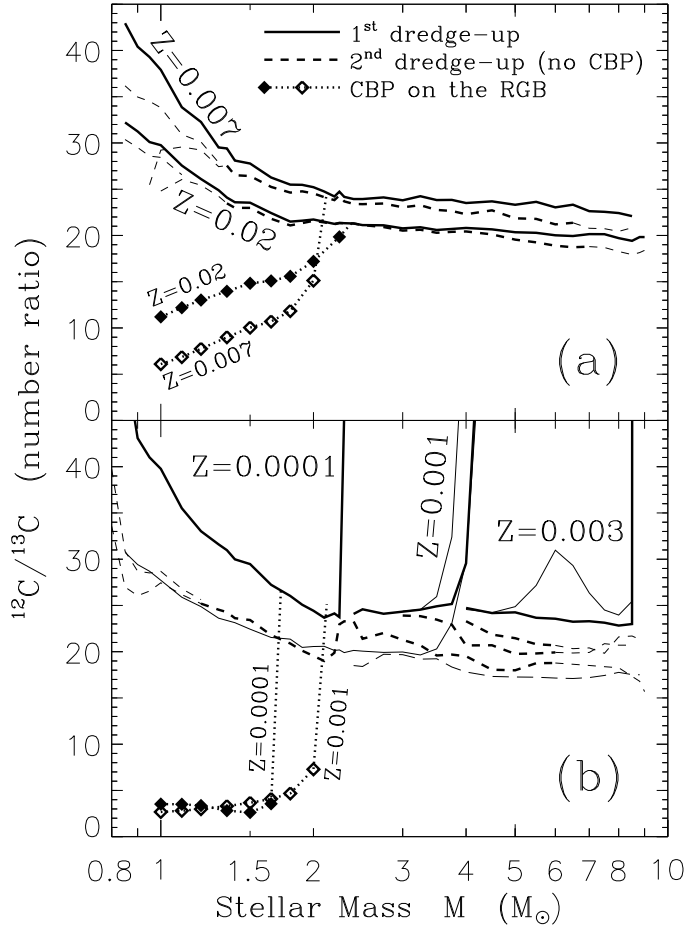}{4.5 true in}{0}{54}{54}{-180}{-43}

\caption{Theoretical \hbox{$\rm{}^{12}C/{}^{13}C$} number ratios for
first dredge-up ({\it heavy solid curves\/})
on the early RGB ({\it light solid curves\/} show the effect of using LAOL
opacities, rather than OPAL), and for second dredge-up ({\it dashed curves\/})
on the early AGB\hbox{}.
(Note that these standard second dredge-up results for low mass stars do
{\it not\/} take into account any composition changes due to earlier
CBP; at low masses, the upper dashed curve shows results without mass loss.)
{\it Diamonds\/} show the effects of CBP on the RGB\hbox{}.
Initial stellar \hbox{$\rm{}^{12}C/{}^{13}C$} ratios were assumed to be
large at low metallicity (see Table~\protect\ref{tblinitxi}).
(a)~For Population~I metallicities ($Z = 0.02$ and~0.007; for clarity,
the $Z = 0.012$ curve, intermediate between these, is omitted).
(b)~For Population~II metallicities ($Z = 0.0001$, 0.001, and~0.003 --- for
clarity, the latter are omitted at low masses, where their first dredge-up
curves all coincide in any case); lower {\it light solid\/} and
{\it long-dashed curves\/} show the results for $Z = 0.001$ of assuming
the initial stellar \hbox{$\rm{}^{12}C/{}^{13}C$} ratio is independent
of metallicity.
\label{figc13}
}

\end{figure}
\placefigure{figc13}

Figure~\ref{figc13} illustrates the effect on \hbox{$\rm{}^{12}C/{}^{13}C$}
ratios of varying the metallicity and the interior opacities.  For the
dredge-up curves, the increasing
trend with metallicity is due to our assumption of
an increasing trend in the initial stellar \hbox{$\rm{}^{12}C/{}^{13}C$}
ratio (see Table~\ref{tblinitxi} and~\S~\ref{methods}).  If one assumes
instead that the initial stellar \hbox{$\rm{}^{12}C/{}^{13}C$} ratio is
independent of metallicity (as Charbonnel [1994]\markcite{Char94} did),
one finds that the dredge-up curves for $Z = 0.001$ (lower light solid
and long-dashed curves in Fig.~\ref{figc13}b) lie slightly
lower than the curves for $Z = 0.02$, in agreement with the results of
Charbonnel (1994)\markcite{Char94}.
Figure~\ref{figc13} also shows the final RGB \hbox{$\rm{}^{12}C/{}^{13}C$}
ratio predicted by our ``evolving RGB'' CBP models ({\it diamonds\/}); the
ratio produced by CBP in Population~II RGB
stars is expected to approach the CN~cycle equilibrium value of~$\sim 3$,
due to their higher H-shell temperatures (see~\S~\ref{strencbp}).


\begin{figure}[!t]

     \bootplotfidtwo{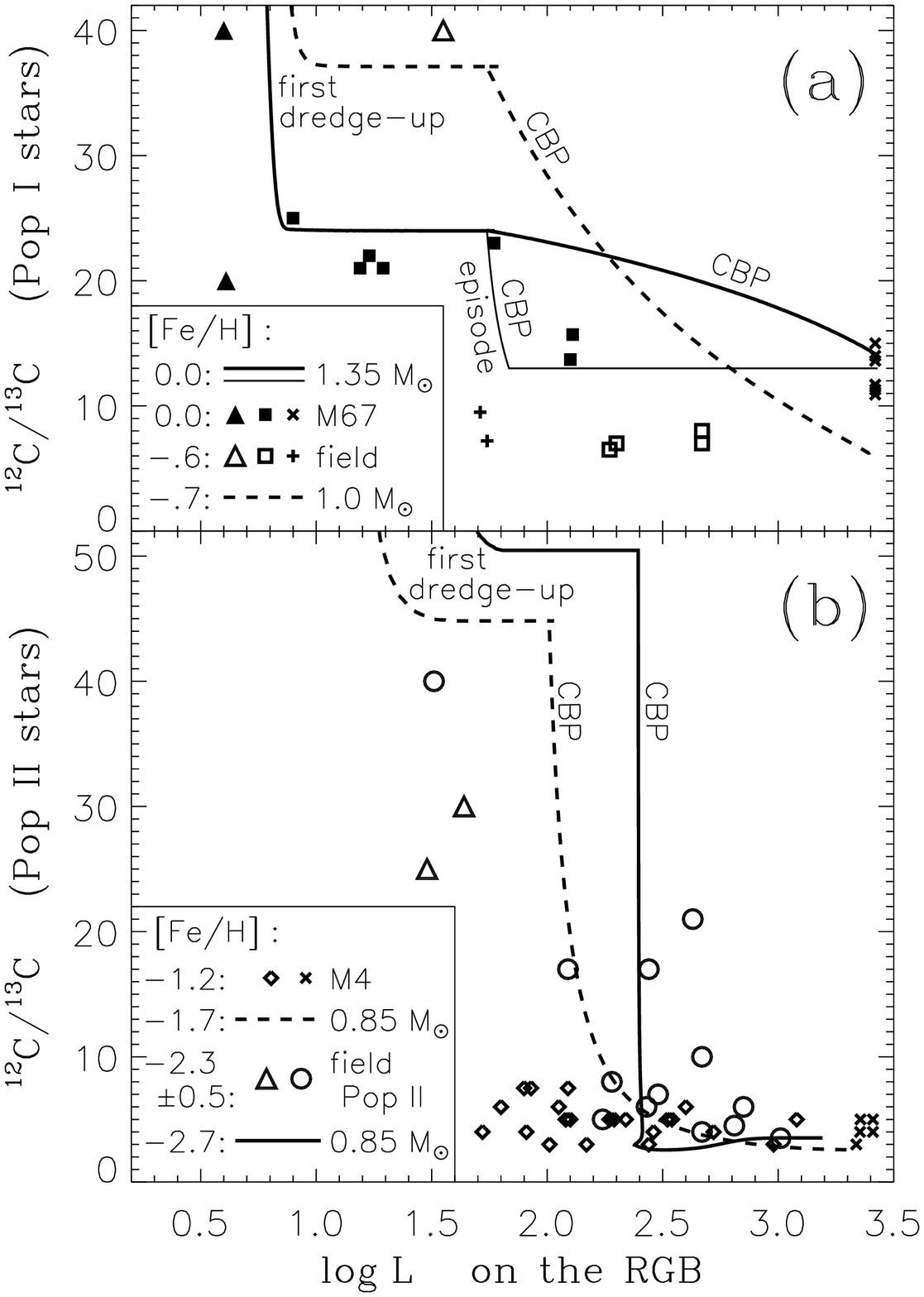}{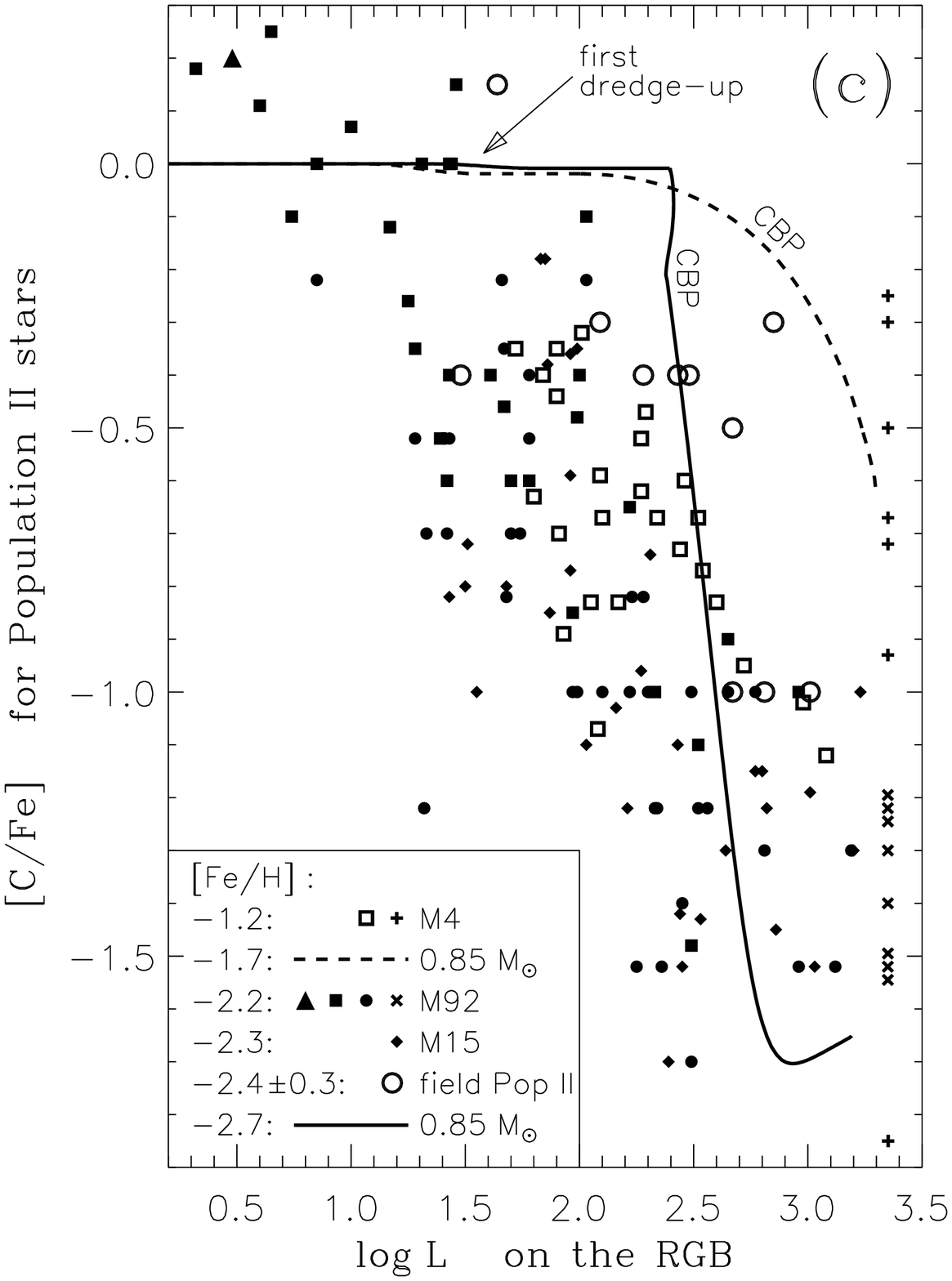}{4 true
       in}{48}{-190}{-40}{48}{-105}{-40}

\caption{Comparison between results of our ``evolving RGB'' CBP models
({\it solid\/} and {\it dotted lines}, as indicated) and observed
\hbox{$\rm{}^{12}C/{}^{13}C$} and [C/Fe] ratios (symbols, as
indicated), as a function of RGB luminosity.  Triangles are lower limits;
symbols ``$+$'' or~``$\times$'' indicate
post-RGB stars, shifted to RGB-tip luminosities for clarity (except for
the two field Population~I stars suspected to be post-RGB stars).
M67:~Gilroy \& Brown (1991)\protect\markcite{GilB91};
field Population~I stars, with luminosities from HIPPARCOS parallaxes:
Charbonnel et al.\ (1998)\protect\markcite{CharBW98};
M4:~Suntzeff \& Smith (1991)\protect\markcite{SuntS91};
field Population~II stars, with relatively uncertain luminosities read
off M92's giant branch at the observed stellar $(B-V)_0$:
Sneden et al.\ (1986)\protect\markcite{SnePV86}.
For~(c), M92:~{\it filled squares\/}
Langer et al.\ (1986)\protect\markcite{Lang+86},
{\it filled circles\/}
Carbon et al.\ (1982)\protect\markcite{Carb+82};
M15:~Trefzger et al.\ (1983)\protect\markcite{Tref+83}.
Bolometric corrections from
VandenBerg (1992)\protect\markcite{Vand92}.
\label{figcbp}
}

\end{figure}
\placefigure{figcbp}

Figure~\ref{figcbp} compares the observed \hbox{$\rm{}^{12}C/{}^{13}C$}
and [C/Fe] ratios as a function of RGB luminosity with the results of
our ``evolving RGB'' CBP models of appropriate mass (lines marked
``CBP'')\hbox{}.  In the open cluster~M67, RGB stars should have masses
of~$\sim 1.3\>M_\odot$ (the turn-off mass being~$\sim 1.2\>M_\odot$:
Gilroy [1989]\markcite{Gil89}).  Comparing the (sparse!) M67 data
(solid symbols in Fig.~\ref{figcbp}a) with our $1.35\>M_\odot$ model
(solid line) suggests that
CBP does indeed begin at the point where the H-burning shell erases
the molecular weight discontinuity (``$\mu$-barrier'') left behind by
first dredge-up, as we had assumed (see~\S~\ref{methods}), but
that CBP begins more strongly than in our ``evolving RGB'' models and
slows down or even stops for luminosities $\log\,L \gtrsim 2.2$ (this
is less extreme than our ``single episode'' CBP case, shown by
the light solid line marked ``CBP episode'' in Fig.~\ref{figcbp}a).
However, the data are too sparse to provide very strong constraints on the
strength of CBP as a function of RGB luminosity --- one cannot completely
rule out even a CBP starting point immediately after first dredge-up (as
appears to occur in globular clusters: see below).
Since M67 was used to normalize our CBP models, they naturally yield
the observed final \hbox{$\rm{}^{12}C/{}^{13}C$} ratio.  The observed field
Population~I RGB stars (open symbols in Fig.~\ref{figcbp}a)
probably have low masses; their observed sub-solar
metallicities ($\rm -0.7 \le [Fe/H] \le -0.42$:
Charbonnel et al.\ 1998\markcite{CharBW98})
suggest an old disk population, and in any case low mass stars are
more common than higher mass stars.  Our $1\>M_\odot$, $Z = 0.007$
``evolving RGB'' CBP model does yield the observed final
\hbox{$\rm{}^{12}C/{}^{13}C$} ratio, but again the observations suggest
CBP begins more strongly and then tails off.

Figure~\ref{figcbp}b presents \hbox{$\rm{}^{12}C/{}^{13}C$} ratios
in Population~II stars.  For {\it field\/} Population~II stars,
our ``evolving RGB'' CBP models are consistent with
the observed \hbox{$\rm{}^{12}C/{}^{13}C$} ratios, though the data are
sparse, and have relatively uncertain luminosities.  For the globular
cluster~M4, the low-luminosity end of the
\hbox{$\rm{}^{12}C/{}^{13}C$} observations suggests that CBP might be
occurring slightly earlier than expected, but the data do not extend far
enough to be certain.
Pilachowski et al.\ (1997)\markcite{Pil+97},
present observations of \hbox{$\rm{}^{12}C/{}^{13}C$} in Population~II
RGB stars as a function of surface gravity, obtaining similar conclusions.
They find that the ratio in field
Population~II stars drops from $\gtrsim 20$ to $< 10$ at $\log\,g \approx 2$
as expected, though the data are again sparse (note deepest first
dredge-up in such stars occurs at $\log\,g \approx 2.5$); they have no data
for low luminosity globular cluster stars ($\log\,g > 2$).  Field
Population~II stars are also observed to undergo further depletions
of~\hbox{$\rm{}^7Li$} (beyond those expected from first dredge-up) at
just the effective temperature where CBP would be expected to begin
(Pilachowski et al.\ 1993\markcite{PilSB93};
see also
Sackmann \& Boothroyd 1998a\markcite{SB98a}).

The [C/Fe] ratio shown in Figure~\ref{figcbp}c is more diagnostic,
especially for globular cluster stars.  The field Population~II star
observations (with metallicities $\rm -2.7 \le [Fe/H] \le -1.8$:
Sneden et al.\ 1986\markcite{SnePV86})
are consistent with our ``evolving RGB'' CBP models.
However, the higher-metallicity globular cluster~M4
($\rm [Fe/H] \approx -1.2$) exhibits {\it more\/} carbon depletion at a
given luminosity than the field stars (rather than less).  Furthermore,
the globular clusters exhibit significant carbon depletion on the RGB
{\it immediately following\/} (or perhaps even before) deepest
{\it first dredge-up\/} on the RGB, {\it long\/} before the H-burning
shell could reach the ``$\mu$-barrier'' that first dredge-up should
create.  For~M4, the relatively high values of [C/Fe] in most of the
post-RGB stars suggests that there may be a ``tail-off'' in the amount
of CBP occurring after a luminosity of $\log\,L \sim 2.5$ is reached
on the RGB, similar to that in Population~I stars.  For the
lower-metallicity clusters M92 and~M15, the [C/Fe] ratio approaches
its CN-cycle equilibrium value near $\log\,L \sim 2.5$ (subsequent CBP
could only increase [C/Fe], if \hbox{$\rm{}^{16}O$} was burned
to~\hbox{$\rm{}^{14}N$}).
Our CBP models find only insignificant burning of~\hbox{$\rm{}^{16}O$},
even at the lowest metallicity we considered, but there are in fact
globular cluster stars
with significant observed O~depletions, as discussed below.

Note that
Charbonnel (1995)\markcite{Char95}
modelled CBP
in 0.8 and~$1.0\>M_\odot$ RGB stars with $Z = 0.001$ and~0.0001; as in our
Population~II models, her reported \hbox{$\rm{}^{12}C/{}^{13}C$} ratios
reached their nuclear equilibrium value shortly after CBP began.
Denissenkov \& Weiss (1996)\markcite{DenW96}
also modelled CBP in a $0.8\>M_\odot$, $Z = 0.0001$ RGB star (assuming that
CBP either began near the base of the RGB or shortly after deepest first
dredge-up), in an attempt to
match the observed trend of \hbox{$\rm{}^{12}C$} depletion with increasing RGB
luminosity.  Their abundance predictions for \hbox{$\rm{}^{12}C$},
\hbox{$\rm{}^{13}C$}, and~\hbox{$\rm{}^{14}N$} are similar to our
Population~II results; like us, they find no depletion
of~\hbox{$\rm{}^{16}O$}.  These models agree with observations of Population~II
field stars and of some globular clusters (e.g., M4, 47~Tuc, NGC~3201,
NGC~2298, NGC~288), which show no O~depletion (see, e.g.,
Kraft 1994\markcite{Kra94},
and references therein).  On the other hand, many globular clusters
show O~depletions on the RGB of as much as an order of magnitude
(e.g., M5, M13, M3, M92, M15, M10, NGC~4833, NGC~362: see
Kraft [1994]\markcite{Kra94}).
Our models do not yield these large O~depletions, when we
normalize by reproducing RGB \hbox{$\rm{}^{12}C/{}^{13}C$} observations
in Population~I stars.  A change in the normalization, i.e., mixing
deeper (to hotter temperatures), can readily produce the observed
O~depletions, as shown by the higher-temperature models of
Denissenkov \& Weiss (1996)\markcite{DenW96};
they found, however, that such models could not simultaneously match the
observed trend of carbon on the RGB
(too much carbon was destroyed).  The calculations of
Denissenkov, Weiss, \& Wagenhuber (1997)\markcite{DenWW97}
show that the observed O~depletions and Al~enhancements cannot be
obtained from primordial contamination of the intracluster medium.
This suggests that there are large star-to-star variations in the depth
of extra mixing in globular clusters, probably due to different stellar
rotation rates; a similar conclusion follows from
the models of
Langer, Hoffmann, \& Sneden (1993)\markcite{LanHS93}
and
Langer \& Hoffman (1995)\markcite{LanH95}.
Kraft (1994)\markcite{Kra94}
discusses evidence that cluster-to-cluster variations are correlated with
differences in average angular momentum and rotation rates.
Star-to-star variation
in the depth of extra mixing is also consistent with the extended
horizontal branches of globular clusters, as pointed out by
Sweigart (1997)\markcite{Swe97}
--- his models indicated that extra mixing deep enough to account for large
O--Na variations could lead to significant conversion of envelope
hydrogen into helium, extending the RGB of such stars and
thus leading to more mass loss and a bluer subsequent horizontal branch
position.  Finally, the fact that a ``bump'' is observed (at the
luminosity predicted by standard models) in the RGB luminosity function
of many clusters that exhibit abundance anomalies suggests that at least
some globular cluster stars do not experience CBP until after the
H-burning shell reaches the ``$\mu$-barrier'' (stars which
experience CBP at an earlier stage would have extra mixing deep
enough to smooth out the composition discontinuity, and would not
spend extra time at the ``bump'' luminosity).


\subsubsection{Nitrogen and Oxygen Isotopes}
 \label{nodredge}

Any envelope carbon depletion is accomplished by converting
the carbon into nitrogen --- the initial C/N ratio (of~$\sim 3.8$,
for the OPAL composition) is reduced by dredge-up to a value of~$\sim 2$
for $1\>M_\odot$ stars.  Dredge-up in intermediate mass stars yields
$\rm C/N \sim 0.7$ for $Z = 0.02$ and $\rm C/N \sim 0.2$ for $Z = 0.0001$,
while CBP in low-mass Population~II stars can reduce carbon to its CN-cycle
equilibrium value of $\rm C/N \sim 10^{-2}$.
Note that hot bottom burning on the thermally pulsing AGB of intermediate
mass stars can also convert carbon into nitrogen, leading to low C/N ratios
(see, e.g.,
Boothroyd et al.\ 1993\markcite{BSA93}).
While \hbox{$\rm{}^{15}N$} was not followed explicitly in the
present work, first and second dredge-up should yield some depletion in
its envelope abundance (somewhat more than that of~\hbox{$\rm{}^{18}O$},
which is discussed below).
Any star that experiences sufficient CBP to significantly affect the
\hbox{$\rm{}^{12}C/{}^{13}C$} ratio should reduce its envelope
\hbox{$\rm{}^{15}N$} abundance to the CN-cycle equilibrium value
(namely, $\hbox{$\rm{}^{14}N/{}^{15}N$} \sim 10^5$), unless the extra
mixing is too slow to process the entire envelope.  (Hot bottom burning in
intermediate mass AGB stars would likewise destroy~\hbox{$\rm{}^{15}N$}.)


\begin{figure}[!t]

     \bootplotfidtwo{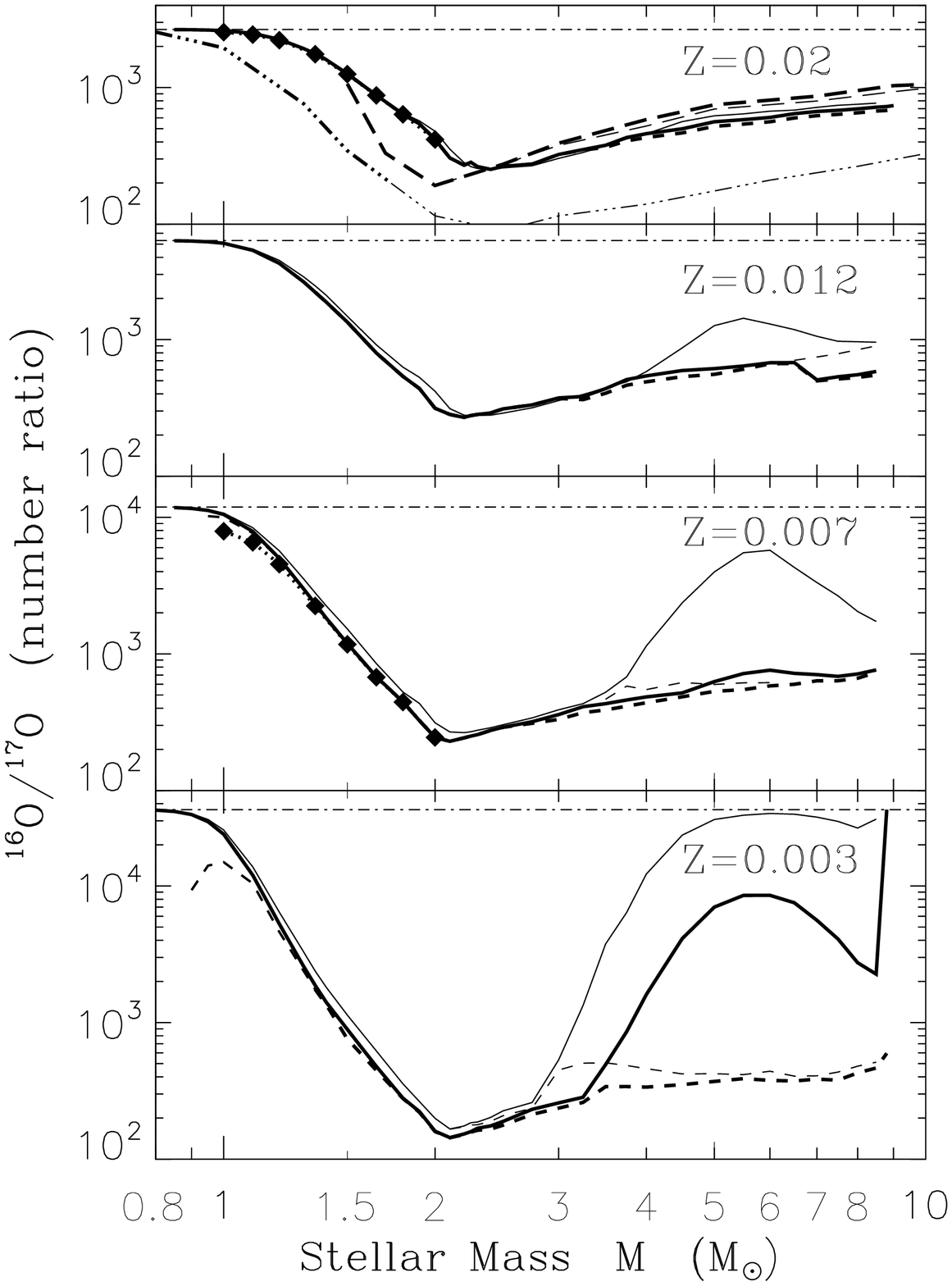}{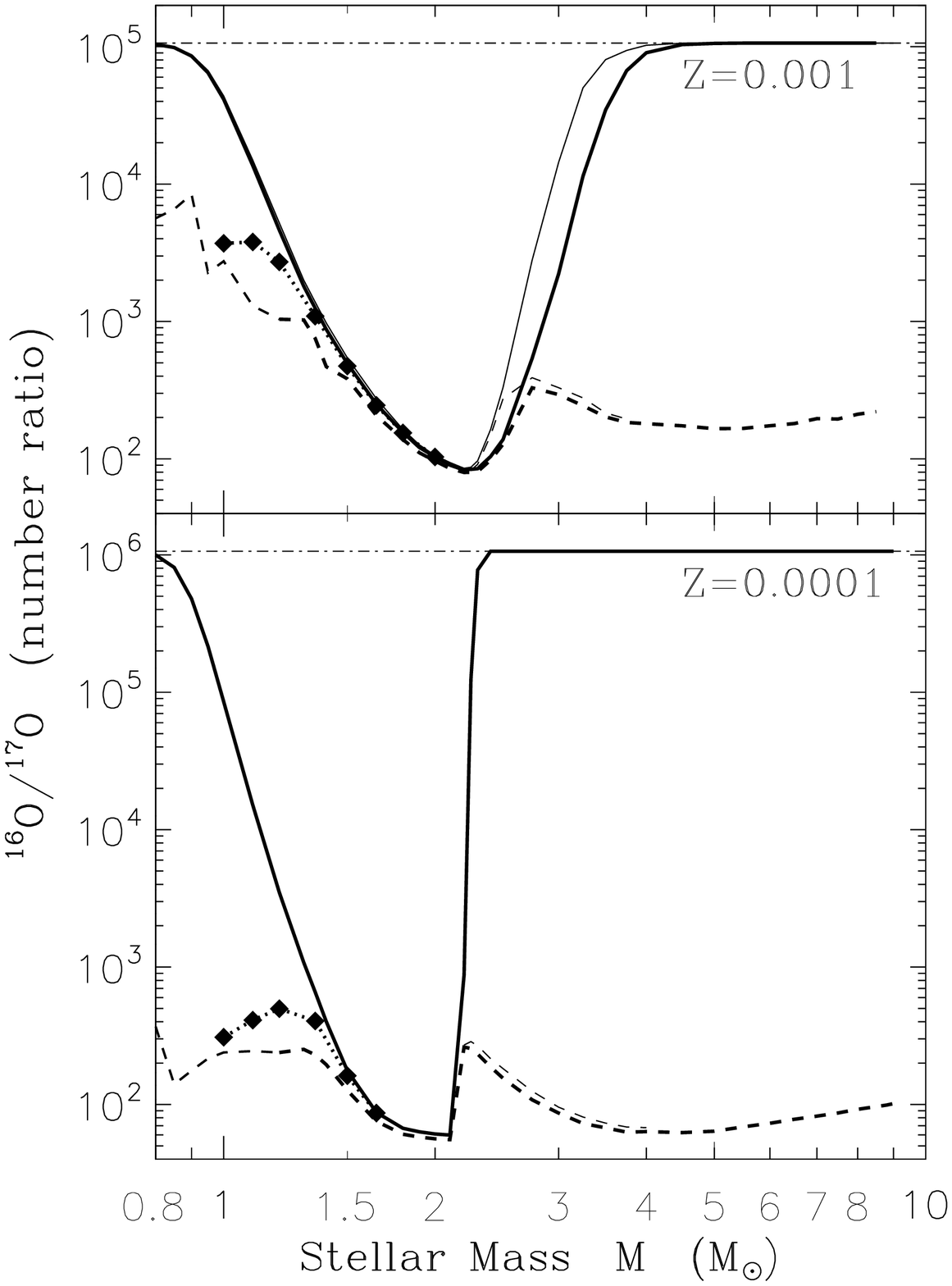}{4 true
       in}{48}{-180}{-40}{48}{-105}{-40}

\caption{Surface \hbox{$\rm{}^{16}O/{}^{17}O$} number ratios from first
dredge-up ({\it solid curves\/}) and second dredge-up ({\it dashed curves\/}),
comparing effect of using OPAL opacities ({\it heavy curves\/}) or the
older LAOL opacities ({\it light curves\/}).  {\it Diamonds\/} show effect
of CBP on the RGB; {\it dot-dashed curves\/} indicate chosen initial stellar
ratios.  For $Z = 0.02$, {\it long-dashed curves\/} indicate results of
Schaller et al.\ (1992)\protect\markcite{Schal+92}, and
{\it triple-dot-dashed curve\/} indicates results of
Dearborn (1992)\protect\markcite{Dea92}.  Note that predicted
\hbox{$\rm{}^{17}O$} enhancements from second dredge-up in
{\it low-mass stars\/} with $Z \le 0.003$ may be spurious (see text).
\label{figo17}
}

\end{figure}
\placefigure{figo17}

Figure~\ref{figo17} illustrates \hbox{$\rm{}^{16}O/{}^{17}O$} ratios,
giving some idea of their uncertainties.  {\it Light curves\/} illustrate
the effect of using the older LAOL interior opacities, rather than the
OPAL opacities ({\it heavy curves\/}).  Although this can make a
significant difference in the point of core helium ignition, and thus
to first dredge-up in intermediate mass stars, the final
\hbox{$\rm{}^{16}O/{}^{17}O$} is less sensitive to the interior opacities,
with agreement generally better than~20\% between the two cases.
The effect of the uncertainty in the $\rm{}^{17}O+p$ reaction rates of
Bl96\markcite{Bl96}
(from the factor $f_1 = 0.31 \pm 0.06$: see~\S~\ref{methods})
is not shown, as it has no effect for low mass stars and leads to an
uncertainty of only $\pm 10$\% for stars with
masses $\gtrsim 2\>M_\odot$; use of the older
La90\markcite{La90}
rates would merely lower the curves by $\sim 20$\% for stars with
masses $\gtrsim 2\>M_\odot$, although the even older (and highly uncertain)
CF88\markcite{CF88}
rates would have yielded large uncertainties for intermediate mass stars
(see~BSW94\markcite{BSW94}).
The effect of assuming a different initial \hbox{$\rm{}^{16}O/{}^{17}O$}
ratio for the low metallicity cases is not shown either, as it has almost
no effect on the part of the graph lying below the chosen initial ratio.

The largest uncertainty in the \hbox{$\rm{}^{16}O/{}^{17}O$} ratios
comes from the fact that dredge-up reaches part-way into a steep
``\hbox{$\rm{}^{17}O$}-peak'', and thus a slight difference in the
depth of dredge-up can have a significant effect on the amount of
\hbox{$\rm{}^{17}O$} that is dredged up (especially in low mass stars).
This is illustrated in the $Z = 0.02$ panel of Figure~\ref{figo17}
by comparing the results of the present work with those of
Schaller et al.\ (1992)\markcite{Schal+92}
({\it long-dashed curves\/}) and of
Dearborn (1992)\markcite{Dea92}
({\it triple-dot-dashed curve\/}).  Above $\sim 2\>M_\odot$, use of the
La90\markcite{La90}
rates by
Schaller et al.\ (1992)\markcite{Schal+92}
would lead one to expect their curve to lie $\sim 20$\% below the curve
of the present work, rather than $\sim 30$\% above it.
Dearborn (1992)\markcite{Dea92}
used the highest allowed rate from the uncertainty range of
CF88\markcite{CF88},
but his curve lies twice as far below the curve of the present work as
one would expect from the difference in nuclear rates alone
(see~BSW94\markcite{BSW94}).
Below $\sim 2\>M_\odot$, different
choices for the $\hbox{$\rm{}^{17}O$}+p$ rates cannot affect the
\hbox{$\rm{}^{16}O/{}^{17}O$} ratio from first dredge-up.
The differences in the predicted \hbox{$\rm{}^{16}O/{}^{17}O$} ratios in low
mass stars must be due to differences in the equation of state, in the way
convection and mass zoning are handled, and (in Dearborn's case) the
opacities, between the different stellar evolution
codes.  The observed \hbox{$\rm{}^{16}O/{}^{17}O$} ratios are not inconsistent
with any of these three theoretical curves, due mainly to the large
uncertainty in the observed stellar masses (see
Dearborn 1992\markcite{Dea92};
El~Eid 1994\markcite{ElE94};
BSW94\markcite{BSW94}).

Significant enhancements of \hbox{$\rm{}^{17}O$} are expected from CBP in
low mass Population~II stars.  Our models suggest that second dredge-up
in low mass Population~II stars also produces large \hbox{$\rm{}^{17}O$}
enhancements (see Fig.~\ref{figo17}); however, it is possible that this
latter may be a spurious effect, due to numerical diffusion from the large,
sharp \hbox{$\rm{}^{17}O$}-peak in the H-burning shell (which must be
rezoned frequently as it burns outwards the early~AGB of these stars).

First and second dredge-up generally cause a slight reduction
($\lesssim 20$\%) in
the surface \hbox{$\rm{}^{18}O$} abundance (see, e.g.,
Dearborn 1992\markcite{Dea92};
El~Eid 1994\markcite{ElE94};
BSW94\markcite{BSW94};
Tables~\ref{tabdrcbp}--\ref{tabdrcbpvlzo} of the present work),
but there is one possible exception.  During second
dredge-up in stars of $\sim 7\>M_\odot$, the
convective envelope may reach into the outer fringes of the region that was
partially mixed by semiconvection during the previous core helium burning
stage, and which thus contains significant amounts of \hbox{$\rm{}^{18}O$}
(produced via the $\hbox{$\rm{}^{14}N$} + \alpha$ reaction); a significant
amount of \hbox{$\rm{}^{18}O$} enrichment can thus result from second
dredge-up in these stars.  In low mass Population~II stars, CBP can yield
large \hbox{$\rm{}^{18}O$} depletions.


\begin{figure}[!t]
  \plotfiddle{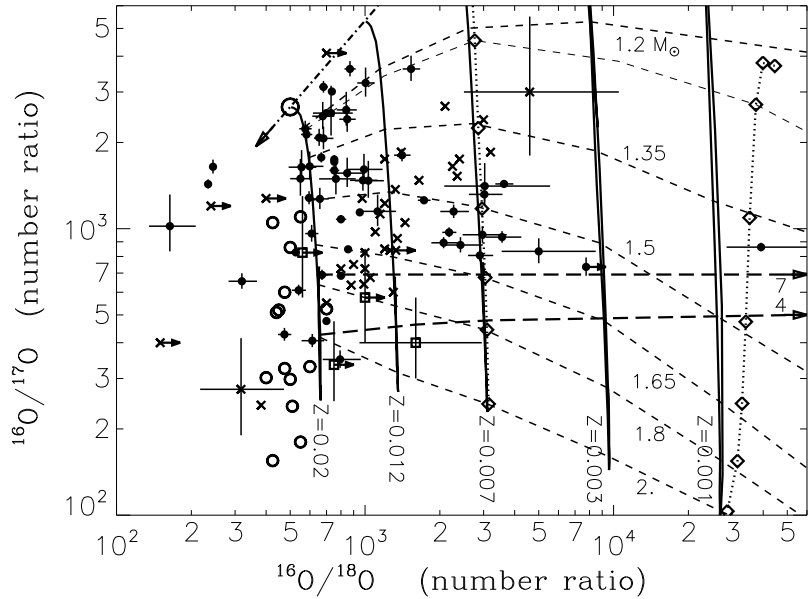}{2.7 true in}{90}{50}{50}{200}{-45}

\caption{Oxygen isotope-isotope plot, comparing theoretical curves with
stellar and grain data.  {\it Dot-dashed line\/} indicates assumed
evolution of the interstellar medium, with \hbox{$\rm{}^{16}O/{}^{17}O$}
and \hbox{$\rm{}^{16}O/{}^{18}O$} inversely proportional to metallicity
(see~\S~\protect\ref{methods}); {\it large open circle\/} indicates the solar
ratios.  {\it Solid lines\/} show isotope ratios resulting from first
dredge-up (full mass range), labelled by metallicity; {\it open diamonds\/}
connected by {\it dotted lines\/}
indicate results of our ``evolving RGB'' CBP models.  {\it Short-dashed
lines\/} connect first dredge-up ratios (for $1.2\>M_\odot$, post-CBP
ratios too) for stars of the same mass (as labelled) but different
metallicities (note that ratios for intermediate mass stars lie roughly
in the same region as for $1.65 - 2\>M_\odot$ stars).  {\it Long-dashed
lines\/} show the effect of hot bottom burning on the TP-AGB for $Z = 0.02$
stars of 4 and~$7\>M_\odot$ (as labelled).  {\it Open circles\/} show
post--dredge-up stellar observations (errorbars omitted) of
Harris et al.\ (1988)\protect\markcite{HarLS88};
{\it open squares\/} show
ratios observed in AGB stars with very low \hbox{$\rm{}^{12}C/{}^{13}C$},
and {\it crosses\/} show ratios of other AGB stars (errorbars omitted,
except for two typical stars, at lower left and at upper right), from
Harris et al.\ (1987)\protect\markcite{Har+87}.
{\it Filled circles\/} show interstellar Al$_2$O$_3$ grain data
measured in meteoritic inclusions by
Nittler et al.\ (1997)\protect\markcite{Nit+97}.
(Rightward-pointing arrows indicate upper limits on observed
\hbox{$\rm{}^{18}O$} abundances.)
\label{figoo}
}

\end{figure}
\placefigure{figoo}

Figure~\ref{figoo} compares theoretical oxygen isotope ratios with
those observed in stars and in interstellar grains; note that most
grains are formed during high-mass-loss episodes in the star's lifetime,
namely, near the tip of the AGB (or, for low mass stars, near
the tip of the RGB\hbox{}).  Note also that, if the oxygen isotope
ratios in the interstellar medium evolve less rapidly than the Fe/H
ratio, the low-metallicity theoretical curves would be shifted to the
left; if the solar oxygen isotope ratios were not typical of the
interstellar medium at solar metallicity, then the theoretical curves
would be shifted as a whole.

The post--dredge-up stellar observations of
Harris et al.\ (1988)\markcite{HarLS88}
({\it open circles\/} in Fig.~\ref{figoo}) lie along a roughly vertical
band consistent with the theoretical predictions
of first dredge-up in stars with initial oxygen isotope ratios very
close to the solar values.  The four AGB stars with
\hbox{$\rm{}^{12}C/{}^{13}C$} near the CN-cycle equilibrium value
({\it open squares\/}) have
oxygen isotope ratios consistent with theoretical predictions for
intermediate mass stars undergoing hot bottom burning (note that three
of these four stars have only an upper limit on \hbox{$\rm{}^{18}O$}).
However, the other AGB stars ({\it crosses\/}) tend to lie along a
band from lower left to upper right, suggesting that low mass AGB
stars may experience \hbox{$\rm{}^{18}O$}~depletion.  Grain data
({\it filled circles\/}) are more precise than stellar data.  While
many of the grain data are consistent with dredge-up in stars of
near-solar metallicity, roughly a dozen grains show
\hbox{$\rm{}^{18}O$}~depletion by factors~$\gtrsim 3$, and several of
of these have \hbox{$\rm{}^{17}O$} abundances a factor of~2 lower than
would be expected in intermediate mass stars undergoing hot bottom
burning.  It is possible, as pointed out by
WBS95\markcite{WBS95},
for CBP on the AGB to account for such \hbox{$\rm{}^{18}O$}~depletion.
However, it would require stronger CBP (deeper extra mixing) on the AGB
than on the RGB ($\Delta\log\,T \approx 0.17$, rather than~0.26).
This might be considered unlikely, considering that CBP appears to grow
weaker as a star climbs the RGB; however, it is possible that
redistribution of angular momentum from a still-rapidly rotating core
during the core helium burning
stage (when the core has expanded and the envelope contracted subsequent
to the RGB, allowing them to couple more strongly) might ``regenerate''
on the AGB the conditions that led to extra mixing and CBP on the RGB
(see, e.g.,
Cohen \& McCarthy 1997\markcite{CohM97};
VandenBerg, Larson, \& de~Propris 1998\markcite{VandLdeP98}).
A more troubling objection is that CBP strong enough to lead to significant
\hbox{$\rm{}^{18}O$}~depletion should also yield a
\hbox{$\rm{}^{12}C/{}^{13}C$} ratio near the CN-cycle equilibrium value
in spite of any \hbox{$\rm{}^{12}C$}~enrichment from third dredge-up
($\rm{}^{12}C/{}^{13}C \lesssim 5$ would be expected, according to
WBS95\markcite{WBS95});
the relatively large \hbox{$\rm{}^{12}C/{}^{13}C$} ratios observed in
most AGB stars suggest that little or no CBP is occurring there.
An episode of strong CBP relatively early on the AGB, followed after
extra mixing tailed off by
third dredge-up of~\hbox{$\rm{}^{12}C$} to increase the
\hbox{$\rm{}^{12}C/{}^{13}C$} ratio to the observed range, cannot
be completely ruled out.  However, it might be considered surprising
that {\it all\/} stars dredged up sufficient
\hbox{$\rm{}^{12}C$} to yield large \hbox{$\rm{}^{12}C/{}^{13}C$} ratios,
even the lowest-mass stars (where third dredge-up
of~\hbox{$\rm{}^{12}C$} is most difficult).  Since CBP yielding the
observed \hbox{$\rm{}^{18}O$}~depletion would destroy only a relatively
small fraction of the envelope carbon (WBS95\markcite{WBS95}), sufficient
third dredge-up to increase \hbox{$\rm{}^{12}C/{}^{13}C$} from $\sim 5$
to~$\sim 20$ would also be sufficient to yield $\rm C/O \sim 1$ --- the
stars observed by
Harris et al.\ (1988)\markcite{HarLS88}
would thus have to be on the verge of becoming carbon stars.  Note that
carbon star observations suggest that stars below a cutoff mass of
$\sim 1.2 - 1.5\>M_\odot$ do not experience sufficient third dredge-up
to become carbon stars (see, e.g.,
Claussen et al.\ 1987\markcite{Clau+87};
Groenewegen \& de~Jong 1993\markcite{GrodeJ93};
Groenewegen, van~den~Hoek, \& de~Jong 1995\markcite{GrovdHdeJ93}).

Stochastic variability in convective mixing and/or interstellar medium
enrichment might yield variations in \hbox{$\rm{}^{17}O$}~abundances from
dredge-up, perhaps allowing the strongly \hbox{$\rm{}^{18}O$}-depleted
grains to be explained by hot bottom burning (though their distribution
in the isotope-isotope plot of Fig.~\ref{figoo} would still be somewhat
surprising).


\subsection{Mixing Events: Location in the H-R Diagram} \label{hrd}


\begin{figure}[!t]
  \plottwo{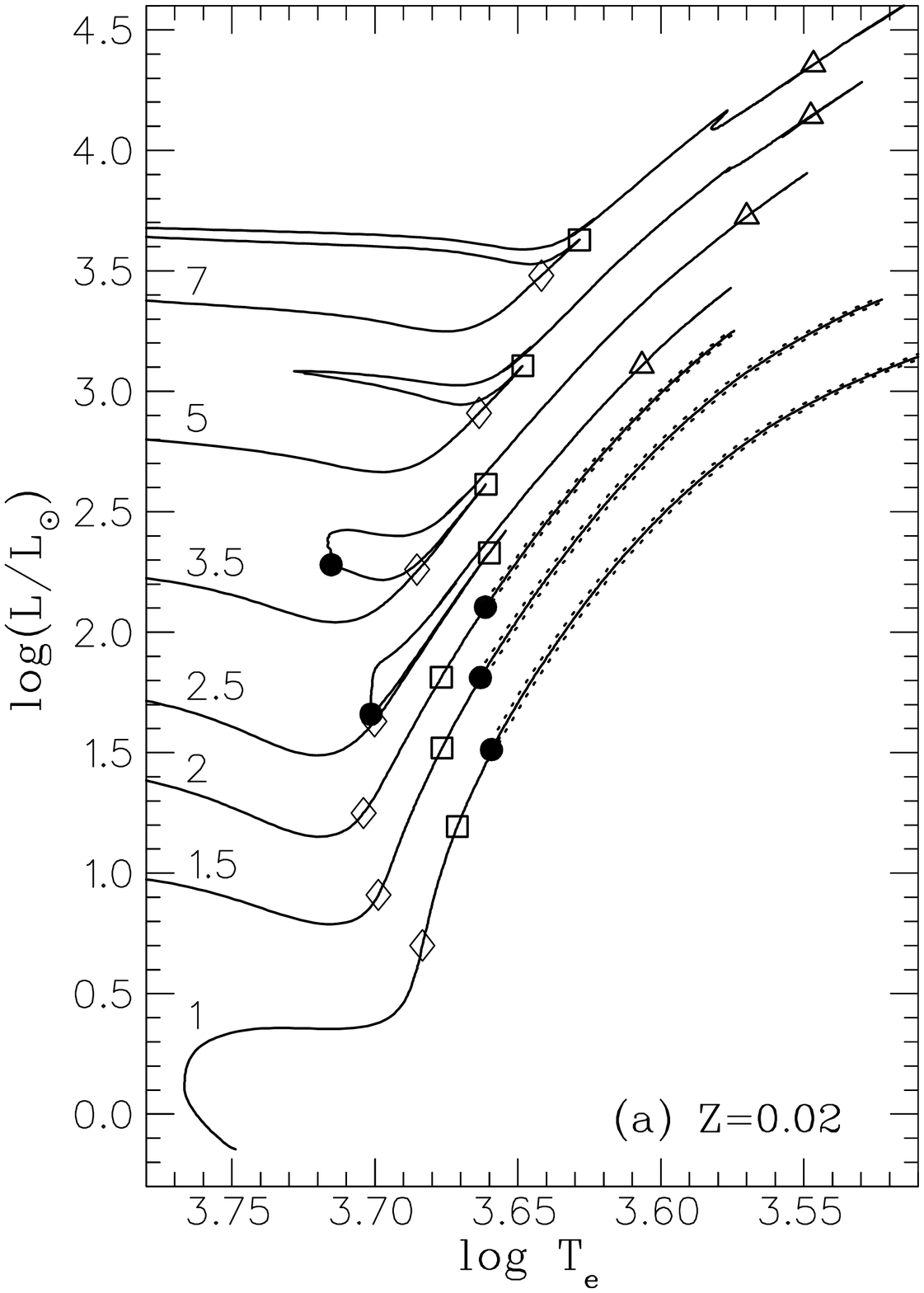}{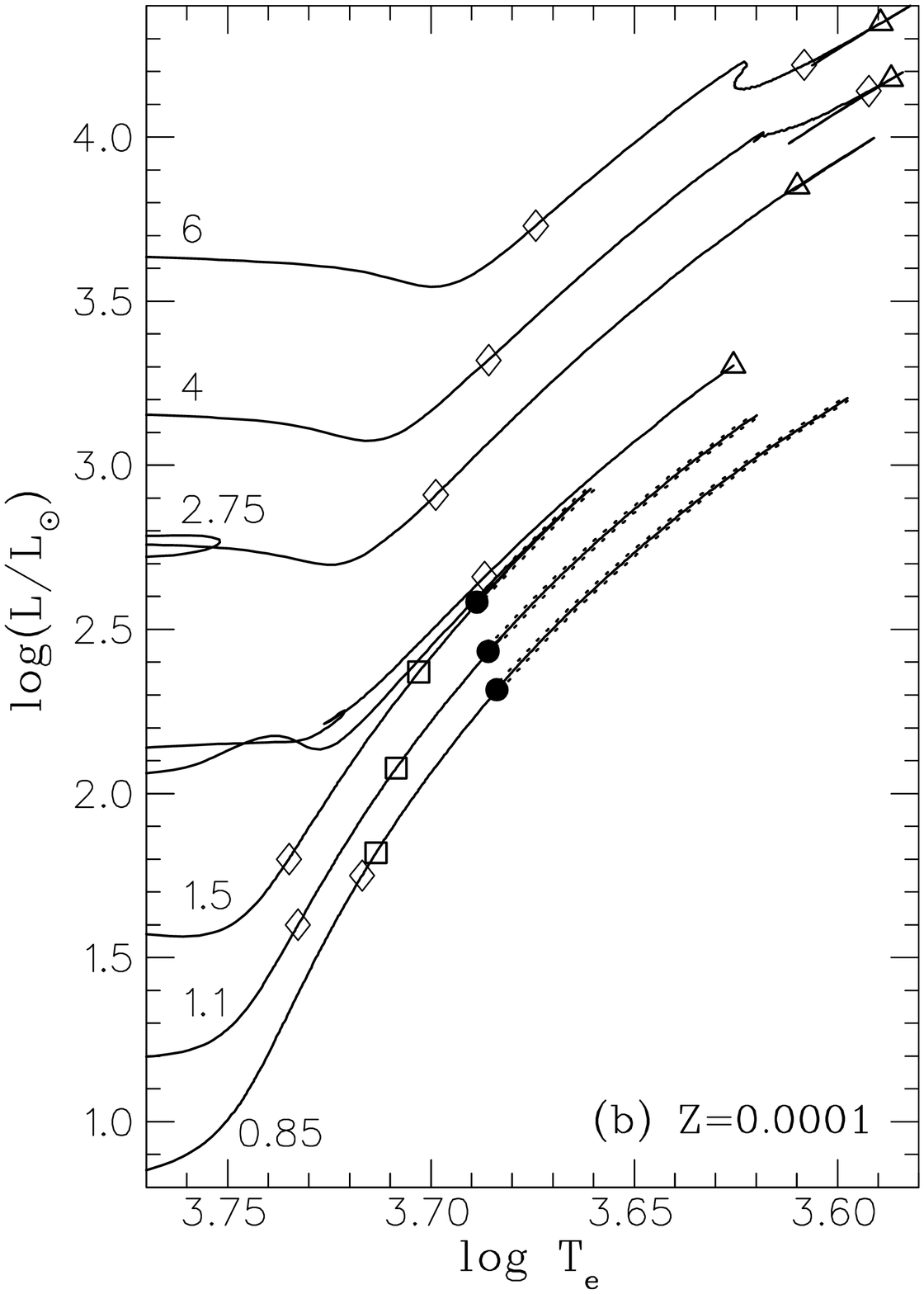}

\caption{The RGB and AGB for (a)~Population~I stars ($Z = 0.02$),
and (b)~Population~II stars ($Z = 0.0001$).  Tracks in the H-R diagram are
labelled by their initial masses; {\it hatched lines\/} indicate
the RGB sections where CBP is expected.  {\it Open
diamonds\/} indicate the points where the convective envelope has engulfed the
entire \hbox{$\rm{}^{13}C$}~pocket (surface \hbox{$\rm{}^{12}C/{}^{13}C$}
ratios are not
changed subsequently by further deepening of the convective envelope).
{\it Open squares\/} on the RGB indicate the end of first dredge-up (the
point on the RGB when the convective envelope reaches deepest into the star).
{\it Solid circles\/} indicate the point where the advancing
hydrogen-burning shell has caught up to (and erased)
the composition discontinuity (``$\mu$-barrier'')
left behind by first dredge-up.  {\it Open
triangles\/} on the AGB indicate the end of second dredge-up (the point
on the AGB when the convective envelope reaches deepest into the star).
For Population~II stars in~(b), dredge-up of the
\hbox{$\rm{}^{13}C$}~pocket ({\it open diamonds\/}) takes place on the
AGB rather than the RGB for stars of $\gtrsim 2.5\>M_\odot$; for
the 4 and~$6\>M_\odot$ cases, the
second {\it open diamond\/} higher on the AGB indicates the end of a
second (less extensive) surface \hbox{$\rm{}^{13}C$} increase.
\label{fighrd}
}

\end{figure}
\placefigure{fighrd}

For stars of solar metallicity, Figure~\ref{fighrd}a shows the
position on the RGB and AGB where a number of important events take place.
One can see that the end of the \hbox{$\rm{}^{13}C$} enrichment phase
({\it diamonds\/}) occurs much earlier on the RGB than the point of deepest
dredge-up ({\it squares\/}: where the convective envelope reaches in
the furthest).  For low mass stars ($M \lesssim 2 \>M_\odot$), both these
events take place fairly early on the~RGB; subsequently, the convective
envelope retreats outwards.  The deepest convective
penetration leaves behind a large composition discontinuity, which is
also a major discontinuity in the mean molecular weight~$\mu$, i.e.,
a ``$\mu$-barrier''; as discussed in \S~\ref{intro}, this
is expected act as a barrier to
any extra mixing that occurs below the base of the convective envelope.
As the hydrogen-burning shell eats its way outwards through the star,
it eventually reaches and destroys this $\mu$-barrier ({\it circles\/}),
allowing CBP to begin.  For $M \gtrsim 2.5\>M_\odot$, one sees from
Figure~\ref{fighrd}a
that deepest first dredge-up occurs near the tip of the RGB, and the
$\mu$-barrier is not destroyed until later.  In these cases, no CBP
is expected on the RGB; this is in agreement with
the fact that no excess \hbox{$\rm{}^{13}C$} enhancements are observed for
$M \gtrsim 2.5\>M_\odot$ (see, e.g.,
Gilroy 1989\markcite{Gil89};
WBS95\markcite{WBS95}).

Let us now consider the AGB\hbox{}.  In principle,
CBP is possible whenever no $\mu$-barrier exists to
prevent it.  For stars of $M \lesssim 3.5\>M_\odot$, the $\mu$-barrier
created by first dredge-up is destroyed prior to the base of the
AGB --- such stars might
experience CBP on the early AGB\hbox{}.
However, the driving mechanism for extra mixing may be weaker or non-existent
by that point (as discussed in \S~\ref{dredge}),
and in any case hydrogen shell burning is weak on the early AGB of
intermediate mass stars (most of the star's luminosity being supplied by the
helium-burning shell), and thus not much CBP is
expected.  After the onset of the helium shell
flashes (thermal pulses), i.e., on the TP-AGB, the hydrogen shell burns
strongly again, and second dredge-up has wiped out any $\mu$-barrier, so
CBP might occur (if the driving mechanism for extra
mixing is still active).  Note however that
stars of $M \gtrsim 4\>M_\odot$ encounter hot bottom burning on the TP-AGB,
wherein the convective envelope actually reaches into hot regions of
the hydrogen-burning shell, and nuclear burning creates a ``$\mu$-barrier''
just below the base of of convection.

For Population~II stars of $Z = 0.0001$, Figure~\ref{fighrd}b
presents the position in the H-R diagram of the key mixing events, and
one might make an analysis similar to that of Population~I stars above,
with similar conclusions.  As noted in \S~{dredge}, this would yield
predictions that might be consistent with observations of field
Population~II stars, but not with globular cluster observations.


\subsection{A Simple Estimate of Cool Bottom Processing} \label{strencbp}

It is possible to make a rough estimate of the total amount of CBP,
as measured by the amount of processing of CNO isotopes.
We must assume that CBP is be independent of the speed (and geometry)
of mixing;
WBS95\markcite{WBS95}
showed that this condition is generally satisfied
for the CNO isotopes.  If we assume that the inner boundary of deep mixing
is at the point where the $\mu$-gradient becomes non-negligible, in the
outer wing of the hydrogen-burning shell, then by definition the rate
of hydrogen burning at the inner boundary of deep mixing is proportional
the rate of hydrogen burning in the hydrogen-burning shell itself.
The latter is given by~$X (d M_c / dt)$, where $X$~is the envelope hydrogen
mass fraction and $M_c$~is the mass of the hydrogen-exhausted
core; since $X$ does not vary much, we can say that the rate of hydrogen
burning is proportional to~$d M_c / dt$.  The mass of CNO elements processed
in the CBP region is proportional to the mass of
{\it hydrogen\/}
burned there, which in turn is proportional to $d M_c / dt$ --- low
metallicity stars must burn hotter to burn the same amount of hydrogen via
the CNO-cycle, and thus experience more extensive CBP\hbox{}.  The rate of
change in the CNO isotope ratios is proportional to the mass rate of CNO
processing divided by the mass of CNO in the envelope, and thus to
$(d M_c / dt) / ( M_{\rm env} Z_{\rm CNO} )$.
Generally, $M_{\rm env}$
and~$Z_{\rm CNO}$ do not vary much as a function of time.
If we assume that the extra mixing does not ``tail off'' as the star
climbs the RGB, then
we can estimate that the total change in the envelope CNO
isotope ratios due to CBP is proportional to
$\Delta M_c / ( M_{\rm env} Z_{\rm CNO} )$.
While this last assumption is somewhat dubious, since extra mixing
{\it does\/} appear to ``tail off'' (see~\S~\ref{cdredge}),
the above formula probably provides a reasonable estimate on the RGB,
except for globular cluster stars (as shown by
the extent to which our ``evolving RGB'' CBP models fit the final
observed isotope ratios).


\begin{figure}[!t]
  \plotfiddle{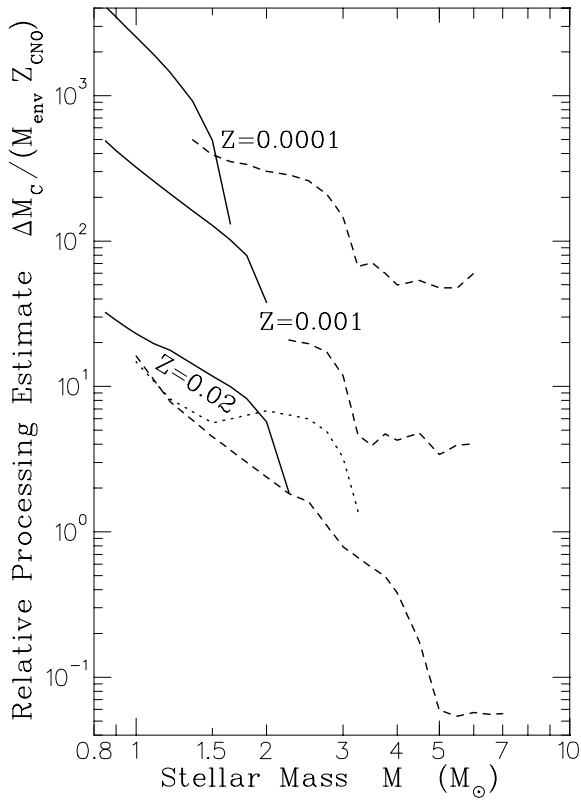}{3.5 true in}{0}{45}{45}{-180}{-36}

\caption{Estimate of relative amounts of CNO CBP
for stars of different initial mass and metallicity, as measured by the
quantity $\Delta M_c / ( M_{\rm env} Z_{\rm CNO} )$: see text.  {\it Solid
lines\/}: for the portion of the RGB where CBP can
take place, i.e., from the point where the molecular weight discontinuity
is erased to the tip of the RGB\hbox{}.  {\it Dashed lines\/}: for the portion
of the early AGB where CBP can take place, up to the point
where helium shell flashes commence.  {\it Dotted line\/}: for the helium shell
flash stage on the AGB (this could only be estimated for the $Z = 0.02$ case).
\label{figestcbp}
}

\end{figure}
\placefigure{figestcbp}

Figure~\ref{figestcbp} displays the value of this CBP estimate
$\Delta M_c / ( M_{\rm env} Z_{\rm CNO} )$ as a function of the stellar mass
for three stages of evolution: the upper RGB ({\it solid\/}), the early AGB
({\it dashed\/}), and the TP-AGB ({\it dotted}, using the Population~I
initial-final mass relation of
Weidemann [1984]\markcite{Weid84}).
Initial-final mass relations for Population~II stars are
too uncertain to be useful for estimates of
the importance of CBP on the TP-AGB of such stars.
We assumed that early AGB processing did not start
until core helium exhaustion, which happens when the star is near the
base of the AGB\hbox{}.  The AGB cases will of course be overestimates
(upper limits), if the driving mechanism for the extra mixing has
at least partially died away prior to the AGB, as seems likely.


\subsection{Interstellar Medium Enrichment} \label{ism}

For stars of $\lesssim 1\>M_\odot$, most of the mass loss takes place
near the tip of the RGB; most of the CBP encountered on the RGB
takes place before much envelope mass is lost.
For stars of $> 1\>M_\odot$, most of the mass loss takes
place even later, during the thermally-pulsing AGB
(Boothroyd \& Sackmann 1988\markcite{BS88};
Sackmann et al.\ 1993\markcite{SBK93};
Boothroyd, Sackmann, \& Ahern 1993\markcite{BSA93});
for stars of solar metallicity, Figure~\ref{figestcbp} shows that even the
upper limit on AGB
CBP is expected to be relatively minor.  A full chemical
evolution model, making use of time delay between star formation and
mass loss on the RGB or AGB
as well as of star formation rates as a function of time and effects of
changing metallicity, is beyond
the scope of this paper; we will make a rough estimate of interstellar
medium enrichment, by considering only solar metallicity, and considering
all stars of mass $> 1\>M_\odot$ (where the time delay until envelope mass
loss does not exceed the age of the galactic disk).

As shown \S~\ref{dredge}, first and second dredge-up result in
large enhancements of \hbox{$\rm{}^{13}C$}, \hbox{$\rm{}^{14}N$},
and~\hbox{$\rm{}^{17}O$} in stellar envelopes, much of which  will
subsequently be injected into the interstellar medium via mass loss,
unless subsequent nucleosynthetic events modify the envelope abundances
prior to the end of the AGB\hbox{}.  Standard stellar models predict that
only minor changes to these isotopes should result from third dredge-up
during the helium shell flash stage on the AGB, but that stars of masses
$\sim 4 - 7\>M_\odot$ should encounter hot bottom burning that can
significantly affect the CNO isotopes.  However, enrichment
of the interstellar medium is not expected to be significantly affected by
hot bottom burning for the isotopes \hbox{$\rm{}^{13}C$} (it
increases \hbox{$\rm{}^{13}C$} relative to \hbox{$\rm{}^{12}C$}, but the
total carbon abundance is generally reduced by an even larger factor)
and~\hbox{$\rm{}^{17}O$} (with the
Bl96\markcite{Bl96}
rates, it yields \hbox{$\rm{}^{16}O/{}^{17}O$} ratios comparable to
those from dredge-up); \hbox{$\rm{}^{18}O$}~destruction in this relatively
narrow mass range yields only a small effect, and \hbox{$\rm{}^{14}N$}
enrichment of the interstellar medium should be increased by only $\sim 20$\%
(Boothroyd et al.\ 1993\markcite{BSA93};
Sackmann \& Boothroyd 1998b\markcite{SB98b}).
In stars of mass $\lesssim 2\>M_\odot$, CBP
can also affect these isotopes, as discussed above in \S~\ref{dredge}.
We shall not specifically consider \hbox{$\rm{}^{12}C$} (which is
largely produced during third dredge-up on the TP-AGB, which we have not
followed), nor~\hbox{$\rm{}^{16}O$} (which is produced by supernovae).

The computations of
Weaver \& Woosley (1993)\markcite{WeaW93}
predict that
supernovae will be a source of some \hbox{$\rm{}^{12}C$}, \hbox{$\rm{}^{13}C$},
and~\hbox{$\rm{}^{14}N$}, and a major source of~\hbox{$\rm{}^{17}O$}
(overproduction factors of 4.0, 2.2, 3.3, and~12.6, respectively, relative to
solar abundances); heavier elements are generally overproduced in their
models by a factor of~$\sim 10$ (as are \hbox{$\rm{}^{16}O$}
and~\hbox{$\rm{}^{18}O$}, overproduced by 9.8 and~16.4, respectively).
The relative importance of these
sources can be estimated by folding the mass of each isotope ejected per
star with the initial mass function~$\phi(M)$, which gives the relative
number of stars formed as a function of their mass~$M$.  The mass
enrichment~$m_i(M)$ of an isotope~$i$ ejected into the interstellar medium
is given by $m_i(M) = M_{\rm ejected}(M) [ X_i(M) - X_{i,\rm ISM} ]$, where
$M_{\rm ejected}(M)$ is the mass of the ejected stellar envelope of a star of
mass~$M$, $X_i(M)$~is the mass fraction of isotope~$i$ in the star's envelope,
and $X_{i,\rm ISM}$ is the mass fraction of isotope~$i$ in the interstellar
medium.  Note that $M_{\rm ejected}(M)$~is the difference
between the initial stellar mass~$M$ and the remnant white dwarf
mass~$M_{\rm wd}$ or neutron star mass~$M_{\rm ns}$.  The value
of~$M_{\rm wd}(M)$ may be obtained from the initial--final mass relationship
(Weidemann \& Koester 1983\markcite{WeidK83};
Weidemann 1984\markcite{Weid84}),
and $M_{\rm ns}(M)$ from the estimates of
Weaver \& Woosley (1993)\markcite{WeaW93},
yielding
\begin{equation}
M_{\rm ejected}(M) = \left\{
\begin{array}{ll}
 M - M_{\rm wd} \approx 0.95 \, M - 0.5\>M_\odot \> , \qquad &
  0.85\>M_\odot < M \le 4\>M_\odot \\
 M - M_{\rm wd} \approx 0.9 \, M - 0.3\>M_\odot \> , \qquad &
  4\>M_\odot < M \lesssim 12\>M_\odot \\
 M - M_{\rm ns} \approx 0.982 \, M - 1.22\>M_\odot \> ,\qquad &
  M \ge 12\>M_\odot \> .
\end{array}
\right. \label{eqmej}
\end{equation}
One may
approximate $\phi(M) \propto M^{-s}$ with $s \sim 2.3$
(Salpeter 1955\markcite{Sal55}).
The mass of isotope~$i$ produced via first and
second dredge-up, relative to that from supernovae, is thus
\begin{equation}
{ m_i({\rm dr}) \over m_i({\rm SN}) } =
  { \int_{M_{\rm lo:dr}}^{M_{\rm hi:dr}} \phi(M) M_{\rm ejected}(M)
    [ X_i(M)_{\rm dr} - X_{i,\rm ISM} ] dM
  \over
    \int_{M_{\rm lo:SN}}^{M_{\rm hi:SN}} \phi(M) M_{\rm ejected}(M)
    [ X_i(M)_{\rm SN} - X_{i,\rm ISM} ] dM
  } \> , \label{eqmi}
\end{equation}
where $M_{\rm lo:dr} \approx 1\>M_\odot$, $M_{\rm hi:dr} \approx M_{\rm lo:SN}
\approx 12\>M_\odot$, and $M_{\rm hi:SN} \sim 40\>M_\odot$.

Performing the above integrals, one finds that dredge-up in low and
intermediate mass stars of solar metallicity
produces about 4.3~times as much \hbox{$\rm{}^{13}C$}
enrichment as supernovae, 2.4~times as much~\hbox{$\rm{}^{14}N$}, and
slightly less~\hbox{$\rm{}^{17}O$}, compared to
Weaver \& Woosley's (1993)\markcite{WeaW93}
supernova models;
if one adds the effect of CBP on the RGB, then the ratio
for \hbox{$\rm{}^{13}C$} is~5.7 (rather than~4.3).
These ratios depend to some extent on the value chosen for the slope
of~$\phi(M)$: a 10\% change in this slope~$s$ results in a $\sim 30$\% change
in the enrichment ratios computed via Equation~(\ref{eqmi}).  As discussed
in~\S~\ref{nodredge}, an uncertainty of perhaps as much as a factor of~2
in the production of \hbox{$\rm{}^{17}O$} may result from the steepness of the
\hbox{$\rm{}^{17}O$} profile into which convection reaches during dredge-up.
Note also that
Weaver \& Woosley (1993)\markcite{WeaW93}
appear to have used the ``recommended''
CF88\markcite{CF88}
rates, which would probably result in
an overestimate of supernova production of~\hbox{$\rm{}^{17}O$} by a
factor of~$\sim 4$ relative to what would be obtained using the newer
Bl96\markcite{Bl96}
(or even
La90\markcite{La90})
rates (see, e.g.,
BSW94\markcite{BSW94}).

For most of the isotopes they consider,
Weaver \& Woosley (1993)\markcite{WeaW93}
find supernova overproduction factors of $\sim 10$ relative to the interstellar
medium abundances.  The total amounts of \hbox{$\rm{}^{13}C$},
\hbox{$\rm{}^{14}N$}, \hbox{$\rm{}^{17}O$}, and~\hbox{$\rm{}^{18}O$}
produced should be consistent with the supernova production of the heavier
elements.  For~\hbox{$\rm{}^{13}C$},
Weaver \& Woosley's (1993)\markcite{WeaW93}
supernova models have an overproduction factor of~2.2; if low and
intermediate mass stars produce an additional amount that is 4.3 to~5.7
times as much, this is equivalent to a supernova overproduction factor
of 12 to~14.  For~\hbox{$\rm{}^{14}N$}, supernovae overproduction is~3.3;
an added amount 2.4~times as much is equivalent to an overproduction factor
of~11, and hot bottom burning would increase this to no more than~14.
For~\hbox{$\rm{}^{18}O$}, the supernova overproduction factor of~16.4 is
only slightly reduced by partial destruction in low and intermediate mass
stars, to 15 or~16.  All three
of these isotopes are thus consistent with the heavier elements.
For~\hbox{$\rm{}^{17}O$},
Weaver \& Woosley (1993)\markcite{WeaW93}
give a supernova overproduction factor of~12.6; if this was an overestimate
by a factor of~4 (as discussed above), their overproduction factor would
be~3.2.  Adding the contribution from low and intermediate mass stars to
this latter value would yield the equivalent
of an overproduction factor of~$\sim 16$; hot bottom burning would
not increase this very much, according to the most recent
Bl96\markcite{Bl96}
\hbox{$\rm{}^{17}O$}-destruction rates
(Sackmann \& Boothroyd 1998b\markcite{SB98b}).
Consistency is thus obtained for \hbox{$\rm{}^{17}O$} with the most recent
rates.  Note that most of the \hbox{$\rm{}^{12}C$}
enrichment is due to third dredge-up in those low to intermediate mass stars
that eventually become carbon stars on the AGB, and essentially all
the~\hbox{$\rm{}^{16}O$} enrichment is due to
supernovae, which produce about the right amount, according to
Weaver \& Woosley (1993)\markcite{WeaW93}.
Thus we have a
reasonably self-consistent picture of the enrichment of the interstellar
medium in CNO elements from stars of near-solar metallicity.

\acknowledgements

We both are indebted to Charles A. Barnes for key support and encouragement
as well as stimulating discussions, and Robert D. McKeown for the support
supplied by the Kellogg Radiation Laboratory.  We wish to express a special
gratitude to Charles W. Peck and Helmut A. Abt for helpful discussions and
support.  We are also indebted to G.~J.~Wasserburg for insightful
discussions and support, and for helpful comments on the manuscript.  One of
us (A.~I.~B.)  wishes to thank Scott D. Tremaine and Peter G. Martin for the
support provided by the Canadian Institute for Theoretical Astrophysics.
This work was supported in part by a grant from the Natural Sciences and
Engineering Research Council of Canada, by a grant from the National
Science Foundation \hbox{PHY 94-20470}, by NASA grant \hbox{NAGW-3337}
to G.~J.~Wasserburg, and by a grant from the Australian Research Council.


\clearpage



\begin{deluxetable}{lrcccccccc}

\tablecaption{Abundances From Dredge-up and CBP for $  Z=0.02$}

\tableheadfrac{.2}
     \tableheadfrac{.08}

\tablehead{
 $M_{init}$ & & \multicolumn{5}{c}{\dotfill mass fractions\dotfill} &
  \multicolumn{3}{c}{\dotfill number ratios\dotfill} \nl
 $(M_\odot)$ & case\tablenotemark{a} & $Y$ & $C/Z$ & $N/Z$ & $O/Z$ &
  \hbox{$\rm{}^3$He} & $\!$\hbox{$\rm{}^{12}C/{}^{13}C$} &
  \hbox{$\rm{}^{16}O/{}^{17}O$}$\!$ & \hbox{$\rm{}^{16}O/{}^{18}O$} }
\startdata    \label{tabdrcbp}
{\bf all} & $\!${\bf Init:} & 0.2800 &   0.1733 &   0.0531 &   0.4823 &
                             8.40($-5$) &     90.0 &     2660 &    500.1 \nl
\tablevspace{3pt}
0.85 &              1st: & 0.3001 &   0.1662 &   0.0617 &   0.4823 &
                               0.001931 &     32.2 &     2656 &    504.5 \nl
\tablevspace{3pt}
1.00 &              1st: & 0.3025 &   0.1530 &   0.0771 &   0.4822 &
                               0.001368 &     29.8 &     2610 &    526.9 \nl
     \tablevspace{-3pt}
     &  $\!\!${\bf CBP:} & 0.3025 &   0.1517 &   0.0794 &   0.4822 &
                             7.28($-5$) &     11.2 &     2544 &    542.7 \nl
     \tablevspace{-3pt}
 & 2nd:\tablenotemark{b} & 0.3025 &   0.1528 &   0.0775 &   0.4822 &
                               0.001156 &     24.8 &     2610 &    527.2 \nl
\tablevspace{3pt}
1.10 &              1st: & 0.3028 &   0.1460 &   0.0853 &   0.4822 &
                               0.001112 &     27.6 &     2491 &    549.1 \nl
     \tablevspace{-3pt}
     &  $\!\!${\bf CBP:} & 0.3028 &   0.1450 &   0.0871 &   0.4822 &
                             8.65($-5$) &     12.2 &     2442 &    562.2 \nl
     \tablevspace{-3pt}
 & 2nd:\tablenotemark{b} & 0.3028 &   0.1459 &   0.0855 &   0.4822 &
                               0.001060 &     26.0 &     2491 &    549.1 \nl
\tablevspace{3pt}
1.20 &              1st: & 0.3015 &   0.1407 &   0.0916 &   0.4821 &
                               0.000923 &     26.1 &     2254 &    567.5 \nl
     \tablevspace{-3pt}
     &  $\!\!${\bf CBP:} & 0.3015 &   0.1399 &   0.0930 &   0.4821 &
                               0.000100 &     13.0 &     2221 &    578.7 \nl
     \tablevspace{-3pt}
 & 2nd:\tablenotemark{b} & 0.3015 &   0.1406 &   0.0917 &   0.4821 &
                               0.000899 &     25.4 &     2254 &    567.8 \nl
\tablevspace{3pt}
1.35 &              1st: & 0.2988 &   0.1334 &   0.1001 &   0.4821 &
                               0.000718 &     24.1 &     1770 &    593.5 \nl
     \tablevspace{-3pt}
     &  $\!\!${\bf CBP:} & 0.2988 &   0.1329 &   0.1010 &   0.4821 &
                               0.000121 &     14.0 &     1754 &    602.4 \nl
     \tablevspace{-3pt}
 & 2nd:\tablenotemark{b} & 0.2988 &   0.1334 &   0.1002 &   0.4821 &
                               0.000704 &     23.5 &     1769 &    593.5 \nl
\tablevspace{3pt}
1.50 &              1st: & 0.2957 &   0.1273 &   0.1073 &   0.4821 &
                               0.000575 &     23.5 &     1260 &    615.4 \nl
     \tablevspace{-3pt}
     &  $\!\!${\bf CBP:} & 0.2957 &   0.1269 &   0.1080 &   0.4821 &
                               0.000136 &     14.8 &     1253 &    623.1 \nl
     \tablevspace{-3pt}
 & 2nd:\tablenotemark{b} & 0.2957 &   0.1272 &   0.1074 &   0.4821 &
                               0.000566 &     23.0 &     1260 &    615.4 \nl
\tablevspace{3pt}
1.65 &              1st: & 0.2930 &   0.1234 &   0.1119 &   0.4821 &
                               0.000467 &     22.4 &    879.0 &    632.8 \nl
     \tablevspace{-3pt}
     &  $\!\!${\bf CBP:} & 0.2930 &   0.1231 &   0.1125 &   0.4821 &
                               0.000141 &     15.1 &    876.0 &    639.5 \nl
     \tablevspace{-3pt}
 & 2nd:\tablenotemark{b} & 0.2931 &   0.1232 &   0.1121 &   0.4821 &
                               0.000461 &     21.8 &    879.0 &    632.8 \nl
\tablevspace{3pt}
1.80 &              1st: & 0.2909 &   0.1202 &   0.1156 &   0.4821 &
                               0.000391 &     21.5 &    637.7 &    646.5 \nl
     \tablevspace{-3pt}
     &  $\!\!${\bf CBP:} & 0.2909 &   0.1199 &   0.1161 &   0.4821 &
                               0.000155 &     15.6 &    636.5 &    652.3 \nl
     \tablevspace{-3pt}
 & 2nd:\tablenotemark{b} & 0.2909 &   0.1200 &   0.1158 &   0.4821 &
                               0.000387 &     21.1 &    637.7 &    646.5 \nl
\tablevspace{3pt}
2.00 &              1st: & 0.2888 &   0.1167 &   0.1197 &   0.4820 &
                               0.000319 &     21.7 &    417.0 &    661.0 \nl
     \tablevspace{-3pt}
     &  $\!\!${\bf CBP:} & 0.2888 &   0.1166 &   0.1200 &   0.4820 &
                               0.000173 &     17.2 &    416.7 &    664.6 \nl
     \tablevspace{-3pt}
 & 2nd:\tablenotemark{b} & 0.2888 &   0.1167 &   0.1197 &   0.4820 &
                               0.000318 &     21.6 &    417.0 &    661.0 \nl
\tablevspace{3pt}
2.20 &              1st: & 0.2892 &   0.1153 &   0.1249 &   0.4781 &
                               0.000267 &     21.3 &    270.4 &    662.3 \nl
     \tablevspace{-3pt}
 & 2nd:\tablenotemark{b} & 0.2892 &   0.1154 &   0.1249 &   0.4781 &
                               0.000267 &     21.2 &    270.4 &    662.3 \nl
\tablevspace{3pt}
2.25 &              1st: & 0.2896 &   0.1155 &   0.1260 &   0.4766 &
                               0.000258 &     21.4 &    283.9 &    662.4 \nl
     \tablevspace{-3pt}
     &  $\!\!${\bf CBP:} & 0.2896 &   0.1154 &   0.1260 &   0.4766 &
                               0.000212 &     19.8 &    283.9 &    663.2 \nl
     \tablevspace{-3pt}
 & 2nd:\tablenotemark{b} & 0.2896 &   0.1155 &   0.1260 &   0.4766 &
                               0.000258 &     21.3 &    283.9 &    662.4 \nl
\tablevspace{3pt}
2.40 &              1st: & 0.2907 &   0.1146 &   0.1300 &   0.4731 &
                               0.000229 &     21.3 &    251.6 &    663.1 \nl
     \tablevspace{-3pt}
     &              2nd: & 0.2907 &   0.1146 &   0.1300 &   0.4731 &
                               0.000229 &     21.3 &    251.6 &    663.1 \nl
\tablevspace{3pt}
2.50 &              1st: & 0.2915 &   0.1143 &   0.1324 &   0.4708 &
                               0.000213 &     21.1 &    264.7 &    661.2 \nl
     \tablevspace{-3pt}
     &              2nd: & 0.2915 &   0.1143 &   0.1324 &   0.4708 &
                               0.000213 &     21.1 &    264.7 &    661.2 \nl
\tablevspace{3pt}
     \tablebreak
2.75 &              1st: & 0.2934 &   0.1134 &   0.1376 &   0.4661 &
                               0.000180 &     21.1 &    275.3 &    658.8 \nl
     \tablevspace{-3pt}
     &              2nd: & 0.2934 &   0.1134 &   0.1376 &   0.4661 &
                               0.000179 &     20.9 &    275.3 &    658.8 \nl
\tablevspace{3pt}
3.00 &              1st: & 0.2944 &   0.1131 &   0.1408 &   0.4628 &
                               0.000156 &     20.8 &    323.6 &    656.6 \nl
     \tablevspace{-3pt}
     &              2nd: & 0.2944 &   0.1126 &   0.1414 &   0.4628 &
                               0.000155 &     20.5 &    323.4 &    658.3 \nl
\tablevspace{3pt}
3.50 &              1st: & 0.2949 &   0.1129 &   0.1441 &   0.4592 &
                               0.000123 &     20.6 &    378.0 &    653.5 \nl
     \tablevspace{-3pt}
     &              2nd: & 0.2949 &   0.1119 &   0.1453 &   0.4593 &
                               0.000122 &     20.3 &    364.8 &    658.1 \nl
\tablevspace{3pt}
4.00 &              1st: & 0.2941 &   0.1131 &   0.1450 &   0.4580 &
                               0.000103 &     20.8 &    459.6 &    652.1 \nl
     \tablevspace{-3pt}
     &              2nd: & 0.2942 &   0.1115 &   0.1472 &   0.4575 &
                               0.000101 &     20.4 &    430.7 &    658.8 \nl
\tablevspace{3pt}
4.50 &              1st: & 0.2936 &   0.1133 &   0.1455 &   0.4571 &
                             8.89($-5$) &     20.7 &    498.8 &    650.5 \nl
     \tablevspace{-3pt}
     &              2nd: & 0.3034 &   0.1100 &   0.1555 &   0.4501 &
                             8.59($-5$) &     20.1 &    465.6 &    658.8 \nl
\tablevspace{3pt}
5.00 &              1st: & 0.2933 &   0.1138 &   0.1457 &   0.4563 &
                             7.93($-5$) &     20.3 &    561.7 &    650.7 \nl
     \tablevspace{-3pt}
     &              2nd: & 0.3197 &   0.1077 &   0.1682 &   0.4386 &
                             7.44($-5$) &     19.5 &    519.9 &    661.2 \nl
\tablevspace{3pt}
5.50 &              1st: & 0.2930 &   0.1137 &   0.1462 &   0.4557 &
                             7.23($-5$) &     20.3 &    582.0 &    650.8 \nl
     \tablevspace{-3pt}
     &              2nd: & 0.3326 &   0.1055 &   0.1786 &   0.4296 &
                             6.62($-5$) &     19.3 &    539.8 &    662.4 \nl
\tablevspace{3pt}
6.00 &              1st: & 0.2934 &   0.1133 &   0.1480 &   0.4543 &
                             6.69($-5$) &     20.0 &    604.3 &    650.9 \nl
     \tablevspace{-3pt}
     &              2nd: & 0.3438 &   0.1034 &   0.1885 &   0.4212 &
                             6.00($-5$) &     18.9 &    561.6 &    664.1 \nl
\tablevspace{3pt}
6.50 &              1st: & 0.2941 &   0.1133 &   0.1494 &   0.4526 &
                             6.28($-5$) &     19.9 &    644.1 &    651.4 \nl
     \tablevspace{-3pt}
     &              2nd: & 0.3531 &   0.1020 &   0.1963 &   0.4140 &
                             5.54($-5$) &     18.7 &    598.4 &    664.9 \nl
\tablevspace{3pt}
7.00 &              1st: & 0.2948 &   0.1123 &   0.1522 &   0.4508 &
                             5.94($-5$) &     20.1 &    668.2 &    652.3 \nl
     \tablevspace{-3pt}
     &              2nd: & 0.3606 &   0.1000 &   0.2039 &   0.4080 &
                             5.17($-5$) &     18.8 &    622.1 &    664.0 \nl
\tablevspace{3pt}
7.50 &              1st: & 0.2962 &   0.1119 &   0.1551 &   0.4481 &
                             5.67($-5$) &     20.1 &    680.8 &    652.5 \nl
     \tablevspace{-3pt}
 & 2nd:\tablenotemark{c} & 0.3673 &   0.0987 &   0.2106 &   0.4020 &
                             4.87($-5$) &     18.7 &    636.2 &    619.1 \nl
\tablevspace{3pt}
8.00 &              1st: & 0.2970 &   0.1118 &   0.1566 &   0.4463 &
                             5.45($-5$) &     19.7 &    698.2 &    653.8 \nl
     \tablevspace{-3pt}
 & 2nd:\tablenotemark{c} & 0.3708 &   0.0983 &   0.2139 &   0.3987 &
                             4.65($-5$) &     18.3 &    653.3 &    518.4 \nl
\tablevspace{3pt}
8.50 &              1st: & 0.2981 &   0.1118 &   0.1587 &   0.4441 &
                             5.27($-5$) &     19.4 &    718.1 &    651.9 \nl
     \tablevspace{-3pt}
 & 2nd:\tablenotemark{c} & 0.3703 &   0.0982 &   0.2152 &   0.3974 &
                             4.50($-5$) &     17.9 &    672.8 &    659.2 \nl
\tablevspace{3pt}
9.00 &              1st: & 0.2998 &   0.1107 &   0.1622 &   0.4415 &
                             5.11($-5$) &     19.8 &    738.1 &    653.6 \nl
     \tablevspace{-3pt}
 & 2nd:\tablenotemark{c} & 0.3232 &   0.1040 &   0.1841 &   0.4254 &
                             4.69($-5$) &     18.4 &    691.0 &    670.4 \nl
\enddata
     \vskip -15pt

\tablenotetext{a}{\ ``Init:''~initial stellar abundances for the models
of this table, ``1st:''~abundances at deepest first dredge-up on the RGB,
``CBP:''~RGB-tip abundances from our ``evolving RGB'' CBP models,
``2nd:''~abundances at deepest second dredge-up on the early~AGB
(from models with no CBP)\hbox{}.  Note power-of-ten notation:
\hbox{8.40($-5$)}${} \equiv 8.40 \times 10^{-5}$.}

\tablenotetext{b}{\ Second dredge-up abundances that {\it would\/}
result if no CBP had taken place on the RGB\hbox{}.}

\tablenotetext{c}{\ Second dredge-up abundances during core
carbon ignition, where the program failed.}

\end{deluxetable}


\begin{deluxetable}{lrcccccccc}

\tablecaption{Abundances From Dredge-up and CBP for $ Z=0.007$}

\tableheadfrac{.2}
     \tableheadfrac{.08}

\tablehead{
 $M_{init}$ & & \multicolumn{5}{c}{\dotfill mass fractions\dotfill} &
  \multicolumn{3}{c}{\dotfill number ratios\dotfill} \nl
 $(M_\odot)$ & case\tablenotemark{a} & $Y$ & $C/Z$ & $N/Z$ & $O/Z$ &
  \hbox{$\rm{}^3$He} & $\!$\hbox{$\rm{}^{12}C/{}^{13}C$} &
  \hbox{$\rm{}^{16}O/{}^{17}O$}$\!$ & \hbox{$\rm{}^{16}O/{}^{18}O$} }
\startdata    \label{tabdrcbpz7o}
{\bf all} & $\!${\bf Init:} & 0.2540 &   0.0952 &   0.0292 &   0.6359 &
                             7.62($-5$) &    402.0 &    11890 &     2233 \nl
\tablevspace{3pt}
0.85 &              1st: & 0.2728 &   0.0909 &   0.0344 &   0.6359 &
                               0.001812 &     42.9 &    11780 &     2257 \nl
\tablevspace{3pt}
0.95 &              1st: & 0.2754 &   0.0850 &   0.0412 &   0.6359 &
                               0.001419 &     39.3 &    11230 &     2330 \nl
\tablevspace{3pt}
1.00 &              1st: & 0.2762 &   0.0825 &   0.0442 &   0.6359 &
                               0.001273 &     37.9 &    10500 &     2386 \nl
     \tablevspace{-3pt}
     &  $\!\!${\bf CBP:} & 0.2762 &   0.0788 &   0.0494 &   0.6359 &
                             2.92($-5$) &     6.09 &     7898 &     2616 \nl
     \tablevspace{-3pt}
 & 2nd:\tablenotemark{b} & 0.2763 &   0.0799 &   0.0473 &   0.6359 &
                               0.001115 &     29.2 &     9985 &     2410 \nl
\tablevspace{3pt}
1.10 &              1st: & 0.2772 &   0.0783 &   0.0492 &   0.6359 &
                               0.001036 &     33.9 &     7880 &     2504 \nl
     \tablevspace{-3pt}
     &  $\!\!${\bf CBP:} & 0.2772 &   0.0757 &   0.0529 &   0.6359 &
                             3.50($-5$) &     6.85 &     6560 &     2699 \nl
     \tablevspace{-3pt}
 & 2nd:\tablenotemark{b} & 0.2772 &   0.0768 &   0.0509 &   0.6359 &
                               0.000963 &     29.6 &     7663 &     2520 \nl
\tablevspace{3pt}
1.20 &              1st: & 0.2764 &   0.0750 &   0.0529 &   0.6360 &
                               0.000868 &     32.3 &     5008 &     2600 \nl
     \tablevspace{-3pt}
     &  $\!\!${\bf CBP:} & 0.2764 &   0.0733 &   0.0556 &   0.6360 &
                             4.25($-5$) &     7.75 &     4537 &     2764 \nl
     \tablevspace{-3pt}
 & 2nd:\tablenotemark{b} & 0.2764 &   0.0740 &   0.0542 &   0.6360 &
                               0.000824 &     29.2 &     4965 &     2614 \nl
\tablevspace{3pt}
1.35 &              1st: & 0.2742 &   0.0708 &   0.0578 &   0.6360 &
                               0.000680 &     29.4 &     2327 &     2727 \nl
     \tablevspace{-3pt}
     &  $\!\!${\bf CBP:} & 0.2742 &   0.0697 &   0.0596 &   0.6360 &
                             5.39($-5$) &     9.00 &     2245 &     2855 \nl
     \tablevspace{-3pt}
 & 2nd:\tablenotemark{b} & 0.2742 &   0.0698 &   0.0591 &   0.6360 &
                               0.000654 &     27.6 &     2311 &     2748 \nl
\tablevspace{3pt}
1.50 &              1st: & 0.2716 &   0.0670 &   0.0624 &   0.6360 &
                               0.000550 &     27.8 &     1192 &     2848 \nl
     \tablevspace{-3pt}
     &  $\!\!${\bf CBP:} & 0.2716 &   0.0662 &   0.0636 &   0.6359 &
                             6.44($-5$) &     10.0 &     1174 &     2957 \nl
     \tablevspace{-3pt}
 & 2nd:\tablenotemark{b} & 0.2717 &   0.0662 &   0.0632 &   0.6360 &
                               0.000534 &     26.4 &     1190 &     2861 \nl
\tablevspace{3pt}
1.65 &              1st: & 0.2690 &   0.0645 &   0.0652 &   0.6359 &
                               0.000452 &     26.3 &    680.3 &     2947 \nl
     \tablevspace{-3pt}
     &  $\!\!${\bf CBP:} & 0.2690 &   0.0639 &   0.0663 &   0.6359 &
                             7.23($-5$) &     10.7 &    675.2 &     3044 \nl
     \tablevspace{-3pt}
 & 2nd:\tablenotemark{b} & 0.2691 &   0.0638 &   0.0661 &   0.6359 &
                               0.000441 &     25.2 &    678.8 &     2966 \nl
\tablevspace{3pt}
1.80 &              1st: & 0.2667 &   0.0627 &   0.0674 &   0.6359 &
                               0.000382 &     25.5 &    445.4 &     3016 \nl
     \tablevspace{-3pt}
     &  $\!\!${\bf CBP:} & 0.2667 &   0.0622 &   0.0682 &   0.6359 &
                             8.46($-5$) &     11.8 &    443.6 &     3096 \nl
     \tablevspace{-3pt}
 & 2nd:\tablenotemark{b} & 0.2667 &   0.0621 &   0.0681 &   0.6359 &
                               0.000374 &     24.6 &    444.9 &     3028 \nl
\tablevspace{3pt}
2.00 &              1st: & 0.2645 &   0.0606 &   0.0701 &   0.6360 &
                               0.000311 &     25.2 &    244.7 &     3098 \nl
     \tablevspace{-3pt}
     &  $\!\!${\bf CBP:} & 0.2645 &   0.0605 &   0.0704 &   0.6360 &
                               0.000117 &     15.1 &    244.5 &     3141 \nl
     \tablevspace{-3pt}
 & 2nd:\tablenotemark{b} & 0.2645 &   0.0603 &   0.0705 &   0.6360 &
                               0.000307 &     24.5 &    244.7 &     3106 \nl
\tablevspace{3pt}
2.10 &              1st: & 0.2642 &   0.0601 &   0.0743 &   0.6319 &
                               0.000284 &     24.8 &    229.1 &     3110 \nl
     \tablevspace{-3pt}
     &              2nd: & 0.2642 &   0.0599 &   0.0746 &   0.6319 &
                               0.000280 &     24.1 &    229.1 &     3114 \nl
\tablevspace{3pt}
2.25 &              1st: & 0.2650 &   0.0592 &   0.0828 &   0.6231 &
                               0.000249 &     24.7 &    250.2 &     3097 \nl
     \tablevspace{-3pt}
     &              2nd: & 0.2650 &   0.0592 &   0.0828 &   0.6231 &
                               0.000248 &     24.3 &    250.2 &     3097 \nl
\tablevspace{3pt}
2.50 &              1st: & 0.2668 &   0.0588 &   0.0931 &   0.6117 &
                               0.000207 &     23.9 &    290.3 &     3088 \nl
     \tablevspace{-3pt}
     &              2nd: & 0.2668 &   0.0584 &   0.0936 &   0.6117 &
                               0.000204 &     23.4 &    290.1 &     3100 \nl
\tablevspace{3pt}
2.75 &              1st: & 0.2669 &   0.0584 &   0.0985 &   0.6062 &
                               0.000176 &     24.1 &    318.6 &     3069 \nl
     \tablevspace{-3pt}
     &              2nd: & 0.2669 &   0.0576 &   0.0995 &   0.6063 &
                               0.000173 &     23.5 &    310.2 &     3099 \nl
\tablevspace{3pt}
     \tablebreak
3.00 &              1st: & 0.2660 &   0.0588 &   0.0997 &   0.6044 &
                               0.000154 &     23.8 &    359.3 &     3057 \nl
     \tablevspace{-3pt}
     &              2nd: & 0.2660 &   0.0577 &   0.1011 &   0.6043 &
                               0.000150 &     23.0 &    331.2 &     3099 \nl
\tablevspace{3pt}
3.50 &              1st: & 0.2616 &   0.0597 &   0.0941 &   0.6094 &
                               0.000124 &     23.8 &    432.1 &     3039 \nl
     \tablevspace{-3pt}
     &              2nd: & 0.2619 &   0.0582 &   0.0976 &   0.6076 &
                               0.000120 &     22.8 &    389.0 &     3095 \nl
\tablevspace{3pt}
3.75 &              1st: & 0.2597 &   0.0602 &   0.0900 &   0.6135 &
                               0.000113 &     23.8 &    459.5 &     3034 \nl
     \tablevspace{-3pt}
     &              2nd: & 0.2609 &   0.0584 &   0.0951 &   0.6102 &
                               0.000109 &     22.6 &    414.0 &     3095 \nl
\tablevspace{3pt}
4.00 &              1st: & 0.2578 &   0.0611 &   0.0841 &   0.6190 &
                               0.000105 &     23.5 &    484.0 &     3018 \nl
     \tablevspace{-3pt}
     &              2nd: & 0.2677 &   0.0588 &   0.0963 &   0.6083 &
                               0.000100 &     22.3 &    439.3 &     3076 \nl
\tablevspace{3pt}
4.50 &              1st: & 0.2554 &   0.0622 &   0.0741 &   0.6292 &
                             9.22($-5$) &     23.7 &    517.9 &     3007 \nl
     \tablevspace{-3pt}
     &              2nd: & 0.2835 &   0.0588 &   0.1097 &   0.5929 &
                             8.66($-5$) &     22.7 &    485.2 &     3018 \nl
\tablevspace{3pt}
5.00 &              1st: & 0.2546 &   0.0635 &   0.0678 &   0.6346 &
                             8.32($-5$) &     23.3 &    626.6 &     2990 \nl
     \tablevspace{-3pt}
     &              2nd: & 0.2999 &   0.0579 &   0.1255 &   0.5761 &
                             7.52($-5$) &     21.9 &    532.0 &     3019 \nl
\tablevspace{3pt}
5.50 &              1st: & 0.2544 &   0.0639 &   0.0661 &   0.6359 &
                             7.60($-5$) &     23.6 &    717.6 &     2970 \nl
     \tablevspace{-3pt}
     &              2nd: & 0.3146 &   0.0566 &   0.1394 &   0.5619 &
                             6.65($-5$) &     21.8 &    547.9 &     3017 \nl
\tablevspace{3pt}
6.00 &              1st: & 0.2543 &   0.0642 &   0.0660 &   0.6357 &
                             7.01($-5$) &     23.0 &    759.9 &     2976 \nl
     \tablevspace{-3pt}
     &              2nd: & 0.3257 &   0.0560 &   0.1494 &   0.5511 &
                             6.02($-5$) &     21.2 &    583.1 &     3005 \nl
\tablevspace{3pt}
6.50 &              1st: & 0.2543 &   0.0637 &   0.0667 &   0.6355 &
                             6.53($-5$) &     23.3 &    720.0 &     2989 \nl
     \tablevspace{-3pt}
     &              2nd: & 0.3344 &   0.0552 &   0.1586 &   0.5416 &
                             5.53($-5$) &     21.4 &    598.5 &     2960 \nl
\tablevspace{3pt}
7.00 &              1st: & 0.2545 &   0.0635 &   0.0701 &   0.6319 &
                             6.13($-5$) &     22.6 &    704.7 &     2992 \nl
     \tablevspace{-3pt}
 & 2nd:\tablenotemark{c} & 0.3413 &   0.0550 &   0.1651 &   0.5345 &
                             5.16($-5$) &     20.8 &    635.5 &     1788 \nl
\tablevspace{3pt}
7.50 &              1st: & 0.2551 &   0.0629 &   0.0757 &   0.6262 &
                             5.79($-5$) &     22.6 &    684.2 &     2994 \nl
     \tablevspace{-3pt}
 & 2nd:\tablenotemark{c} & 0.3439 &   0.0548 &   0.1687 &   0.5307 &
                             4.89($-5$) &     20.7 &    635.7 &     1396 \nl
\tablevspace{3pt}
8.00 &              1st: & 0.2573 &   0.0621 &   0.0865 &   0.6150 &
                             5.48($-5$) &     22.4 &    712.6 &     2995 \nl
     \tablevspace{-3pt}
 & 2nd:\tablenotemark{c} & 0.3412 &   0.0547 &   0.1691 &   0.5302 &
                             4.68($-5$) &     20.4 &    663.0 &     2874 \nl
\tablevspace{3pt}
8.50 &              1st: & 0.2607 &   0.0612 &   0.0977 &   0.6034 &
                             5.21($-5$) &     22.1 &    762.2 &     2987 \nl
     \tablevspace{-3pt}
 & 2nd:\tablenotemark{c} & 0.2718 &   0.0591 &   0.1181 &   0.5829 &
                             4.95($-5$) &     20.8 &    731.6 &     3001 \nl
\enddata
     \vskip -15pt

\tablenotetext{a,b,c}{\ As in Table~\protect\ref{tabdrcbp}.}

\end{deluxetable}


\begin{deluxetable}{lrcccccccc}

\tablecaption{Abundances From Dredge-up and CBP for $ Z=0.001$}

\tableheadfrac{.2}
     \tableheadfrac{.08}

\tablehead{
 $M_{init}$ & & \multicolumn{5}{c}{\dotfill mass fractions\dotfill} &
  \multicolumn{3}{c}{\dotfill number ratios\dotfill} \nl
 $(M_\odot)$ & case\tablenotemark{a} & $Y$ & $C/Z$ & $N/Z$ & $O/Z$ &
  \hbox{$\rm{}^3$He} & $\!$\hbox{$\rm{}^{12}C/{}^{13}C$} &
  \hbox{$\rm{}^{16}O/{}^{17}O$}$\!$ & \hbox{$\rm{}^{16}O/{}^{18}O$} }
\startdata    \label{tabdrcbplzo}
{\bf all} & $\!${\bf Init:} & 0.2400 &   0.0765 &   0.0234 &   0.6728 &
                             7.20($-5$) &     3583 &   106200 &    19910 \nl
\tablevspace{3pt}
0.80 &              1st: & 0.2535 &   0.0755 &   0.0247 &   0.6728 &
                               0.001773 &     56.0 &   104100 &    19980 \nl
\tablevspace{3pt}
0.85 &              1st: & 0.2554 &   0.0730 &   0.0276 &   0.6728 &
                               0.001560 &     44.7 &    98560 &    20110 \nl
\tablevspace{3pt}
0.90 &              1st: & 0.2571 &   0.0702 &   0.0308 &   0.6728 &
                               0.001374 &     42.6 &    85440 &    20410 \nl
\tablevspace{3pt}
1.00 &              1st: & 0.2598 &   0.0654 &   0.0364 &   0.6729 &
                               0.001102 &     39.2 &    41370 &    21500 \nl
     \tablevspace{-3pt}
     &  $\!\!${\bf CBP:} & 0.2598 &   0.0295 &   0.0790 &   0.6728 &
                             4.89($-6$) &     2.67 &     3701 &    44530 \nl
     \tablevspace{-3pt}
 & 2nd:\tablenotemark{b} & 0.2600 &   0.0597 &   0.0433 &   0.6730 &
                               0.000918 &     30.0 &     2745 &    22950 \nl
\tablevspace{3pt}
1.10 &              1st: & 0.2615 &   0.0617 &   0.0408 &   0.6729 &
                               0.000902 &     34.7 &    13630 &    22720 \nl
     \tablevspace{-3pt}
     &  $\!\!${\bf CBP:} & 0.2615 &   0.0348 &   0.0729 &   0.6729 &
                             6.43($-6$) &     2.81 &     3791 &    39900 \nl
     \tablevspace{-3pt}
 & 2nd:\tablenotemark{b} & 0.2617 &   0.0572 &   0.0462 &   0.6733 &
                               0.000784 &     28.6 &     1298 &    23900 \nl
\tablevspace{3pt}
1.20 &              1st: & 0.2625 &   0.0583 &   0.0448 &   0.6729 &
                               0.000756 &     32.9 &     4626 &    23820 \nl
     \tablevspace{-3pt}
     &  $\!\!${\bf CBP:} & 0.2625 &   0.0381 &   0.0690 &   0.6729 &
                             8.21($-6$) &     2.98 &     2711 &    37380 \nl
     \tablevspace{-3pt}
 & 2nd:\tablenotemark{b} & 0.2626 &   0.0547 &   0.0491 &   0.6732 &
                               0.000674 &     28.3 &     1038 &    24820 \nl
\tablevspace{3pt}
1.35 &              1st: & 0.2624 &   0.0546 &   0.0491 &   0.6729 &
                               0.000593 &     30.5 &     1250 &    25140 \nl
     \tablevspace{-3pt}
     &  $\!\!${\bf CBP:} & 0.2624 &   0.0411 &   0.0656 &   0.6729 &
                             1.12($-5$) &     3.31 &     1092 &    35300 \nl
     \tablevspace{-3pt}
 & 2nd:\tablenotemark{b} & 0.2625 &   0.0519 &   0.0524 &   0.6732 &
                               0.000543 &     27.1 &    753.6 &    25980 \nl
\tablevspace{3pt}
1.50 &              1st: & 0.2606 &   0.0518 &   0.0524 &   0.6729 &
                               0.000483 &     28.5 &    495.0 &    26050 \nl
     \tablevspace{-3pt}
     &  $\!\!${\bf CBP:} & 0.2606 &   0.0424 &   0.0641 &   0.6728 &
                             1.44($-5$) &     3.67 &    473.4 &    34050 \nl
     \tablevspace{-3pt}
 & 2nd:\tablenotemark{b} & 0.2607 &   0.0494 &   0.0553 &   0.6732 &
                               0.000447 &     25.7 &    382.4 &    26810 \nl
\tablevspace{3pt}
1.65 &              1st: & 0.2587 &   0.0493 &   0.0554 &   0.6730 &
                               0.000405 &     27.2 &    250.0 &    26580 \nl
     \tablevspace{-3pt}
     &  $\!\!${\bf CBP:} & 0.2587 &   0.0428 &   0.0637 &   0.6729 &
                             1.84($-5$) &     4.09 &    245.4 &    32880 \nl
     \tablevspace{-3pt}
 & 2nd:\tablenotemark{b} & 0.2588 &   0.0470 &   0.0585 &   0.6730 &
                               0.000375 &     24.8 &    207.0 &    27290 \nl
\tablevspace{3pt}
1.80 &              1st: & 0.2569 &   0.0469 &   0.0584 &   0.6727 &
                               0.000342 &     26.5 &    155.9 &    26830 \nl
     \tablevspace{-3pt}
     &  $\!\!${\bf CBP:} & 0.2569 &   0.0428 &   0.0638 &   0.6727 &
                             2.38($-5$) &     4.68 &    154.5 &    31520 \nl
     \tablevspace{-3pt}
 & 2nd:\tablenotemark{b} & 0.2570 &   0.0447 &   0.0618 &   0.6722 &
                               0.000318 &     24.2 &    138.3 &    27520 \nl
\tablevspace{3pt}
2.00 &              1st: & 0.2545 &   0.0449 &   0.0623 &   0.6710 &
                               0.000280 &     25.4 &    103.5 &    26790 \nl
     \tablevspace{-3pt}
     &  $\!\!${\bf CBP:} & 0.2545 &   0.0437 &   0.0640 &   0.6711 &
                             4.66($-5$) &     7.28 &    103.2 &    28710 \nl
     \tablevspace{-3pt}
 & 2nd:\tablenotemark{b} & 0.2547 &   0.0427 &   0.0666 &   0.6693 &
                               0.000261 &     23.1 &     96.5 &    27470 \nl
\tablevspace{3pt}
2.20 &              1st: & 0.2526 &   0.0435 &   0.0674 &   0.6672 &
                               0.000236 &     24.4 &     83.6 &    26550 \nl
     \tablevspace{-3pt}
     &              2nd: & 0.2528 &   0.0413 &   0.0727 &   0.6643 &
                               0.000220 &     22.0 &     79.6 &    27240 \nl
\tablevspace{3pt}
2.30 &              1st: & 0.2507 &   0.0434 &   0.0652 &   0.6698 &
                               0.000220 &     24.7 &     84.8 &    26730 \nl
     \tablevspace{-3pt}
     &              2nd: & 0.2508 &   0.0411 &   0.0712 &   0.6662 &
                               0.000205 &     22.2 &     80.7 &    27460 \nl
\tablevspace{3pt}
2.50 &              1st: & 0.2469 &   0.0448 &   0.0612 &   0.6724 &
                               0.000200 &     24.6 &    138.3 &    27450 \nl
     \tablevspace{-3pt}
     &              2nd: & 0.2472 &   0.0422 &   0.0691 &   0.6671 &
                               0.000184 &     22.0 &    127.9 &    28310 \nl
\tablevspace{3pt}
     \tablebreak
2.75 &              1st: & 0.2435 &   0.0482 &   0.0566 &   0.6729 &
                               0.000184 &     24.1 &    538.1 &    27270 \nl
     \tablevspace{-3pt}
     &              2nd: & 0.2448 &   0.0453 &   0.0667 &   0.6653 &
                               0.000169 &     21.6 &    329.1 &    27950 \nl
\tablevspace{3pt}
3.00 &              1st: & 0.2419 &   0.0521 &   0.0520 &   0.6729 &
                               0.000172 &     24.3 &     2219 &    25590 \nl
     \tablevspace{-3pt}
     &              2nd: & 0.2447 &   0.0454 &   0.0683 &   0.6633 &
                               0.000155 &     23.9 &    293.4 &    27990 \nl
\tablevspace{3pt}
3.25 &              1st: & 0.2409 &   0.0578 &   0.0454 &   0.6728 &
                               0.000166 &     24.5 &    11430 &    23110 \nl
     \tablevspace{-3pt}
     &              2nd: & 0.2489 &   0.0443 &   0.0746 &   0.6578 &
                               0.000135 &     23.8 &    243.6 &    28220 \nl
\tablevspace{3pt}
3.50 &              1st: & 0.2405 &   0.0636 &   0.0387 &   0.6729 &
                               0.000159 &     24.9 &    34700 &    21270 \nl
     \tablevspace{-3pt}
     &              2nd: & 0.2611 &   0.0433 &   0.0861 &   0.6462 &
                               0.000119 &     23.5 &    201.9 &    28110 \nl
\tablevspace{3pt}
3.75 &              1st: & 0.2403 &   0.0692 &   0.0321 &   0.6728 &
                               0.000149 &     25.2 &    66930 &    20430 \nl
     \tablevspace{-3pt}
     &              2nd: & 0.2731 &   0.0429 &   0.0979 &   0.6337 &
                               0.000105 &     22.9 &    184.0 &    28040 \nl
\tablevspace{3pt}
4.00 &              1st: & 0.2402 &   0.0742 &   0.0263 &   0.6728 &
                               0.000140 &     29.6 &    90320 &    20090 \nl
     \tablevspace{-3pt}
     &              2nd: & 0.2831 &   0.0421 &   0.1079 &   0.6236 &
                             9.47($-5$) &     20.7 &    180.3 &    28050 \nl
\tablevspace{3pt}
4.50 &              1st: & 0.2401 &   0.0764 &   0.0236 &   0.6728 &
                               0.000122 &     78.9 &   103600 &    19940 \nl
     \tablevspace{-3pt}
     &              2nd: & 0.3004 &   0.0411 &   0.1256 &   0.6038 &
                             7.94($-5$) &     20.7 &    174.0 &    27950 \nl
\tablevspace{3pt}
5.00 &              1st: & 0.2400 &   0.0765 &   0.0235 &   0.6728 &
                               0.000107 &    255.0 &   105700 &    19920 \nl
     \tablevspace{-3pt}
     &              2nd: & 0.3130 &   0.0405 &   0.1405 &   0.5879 &
                             6.82($-5$) &     19.8 &    165.7 &    27220 \nl
\tablevspace{3pt}
5.50 &              1st: & 0.2400 &   0.0765 &   0.0234 &   0.6728 &
                             9.46($-5$) &    736.4 &   106100 &    19910 \nl
     \tablevspace{-3pt}
     &              2nd: & 0.3235 &   0.0400 &   0.1528 &   0.5745 &
                             6.03($-5$) &     19.8 &    166.7 &    24040 \nl
\tablevspace{3pt}
6.00 &              1st: & 0.2400 &   0.0765 &   0.0234 &   0.6728 &
                             8.49($-5$) &     1853 &   106200 &    19910 \nl
     \tablevspace{-3pt}
     &              2nd: & 0.3312 &   0.0398 &   0.1628 &   0.5639 &
                             5.43($-5$) &     19.9 &    174.1 &    16740 \nl
\tablevspace{3pt}
6.50 &              1st: & 0.2400 &   0.0765 &   0.0234 &   0.6728 &
                             7.20($-5$) &     3583 &   106200 &    19910 \nl
     \tablevspace{-3pt}
 & 2nd:\tablenotemark{c} & 0.3367 &   0.0400 &   0.1705 &   0.5551 &
                             4.96($-5$) &     19.9 &    180.6 &     2535 \nl
\tablevspace{3pt}
7.00 &              1st: & 0.2400 &   0.0765 &   0.0234 &   0.6728 &
                             7.20($-5$) &     3583 &   106200 &    19910 \nl
     \tablevspace{-3pt}
 & 2nd:\tablenotemark{c} & 0.3376 &   0.0397 &   0.1737 &   0.5516 &
                             4.63($-5$) &     19.9 &    196.3 &     3842 \nl
\tablevspace{3pt}
7.50 &              1st: & 0.2400 &   0.0765 &   0.0234 &   0.6728 &
                             7.20($-5$) &     3583 &   106200 &    19910 \nl
     \tablevspace{-3pt}
 & 2nd:\tablenotemark{c} & 0.2487 &   0.0421 &   0.1107 &   0.6199 &
                             4.96($-5$) &     20.9 &    194.9 &    28160 \nl
\tablevspace{3pt}
8.00 &              1st: & 0.2400 &   0.0765 &   0.0234 &   0.6728 &
                             7.20($-5$) &     3583 &   106200 &    19910 \nl
     \tablevspace{-3pt}
 & 2nd:\tablenotemark{c} & 0.2482 &   0.0421 &   0.1128 &   0.6174 &
                             4.73($-5$) &     20.7 &    211.2 &    28230 \nl
\tablevspace{3pt}
8.50 &              1st: & 0.2400 &   0.0765 &   0.0234 &   0.6728 &
                             7.20($-5$) &     3583 &   106200 &    19910 \nl
     \tablevspace{-3pt}
 & 2nd:\tablenotemark{c} & 0.3474 &   0.0598 &   0.1868 &   0.5368 &
                             3.92($-5$) &     29.8 &    220.6 &    227.0 \nl
\enddata
     \vskip -15pt

\tablenotetext{a,c}{\ As in Table~\protect\ref{tabdrcbp}.}

\tablenotetext{b}{\ Second dredge-up abundances that {\it would\/}
result if no CBP had taken place on the RGB; note
\hbox{$\rm{}^{16}O/{}^{17}O$} may be inaccurate (see text).}

\end{deluxetable}


\begin{deluxetable}{lrcccccccc}

\tablecaption{Abundances From Dredge-up and CBP for $Z=0.0001$}

\tableheadfrac{.2}
     \tableheadfrac{.08}

\tablehead{
 $M_{init}$ & & \multicolumn{5}{c}{\dotfill mass fractions\dotfill} &
  \multicolumn{3}{c}{\dotfill number ratios\dotfill} \nl
 $(M_\odot)$ & case\tablenotemark{a} & $Y$ & $C/Z$ & $N/Z$ & $O/Z$ &
  \hbox{$\rm{}^3$He} & $\!$\hbox{$\rm{}^{12}C/{}^{13}C$} &
  \hbox{$\rm{}^{16}O/{}^{17}O$}$\!$ & \hbox{$\rm{}^{16}O/{}^{18}O$} }
\startdata    \label{tabdrcbpvlzo}
{\bf all} & $\!${\bf Init:} & 0.2380 &   0.0765 &   0.0234 &   0.6728 &
                             7.14($-5$) &    35820 &  1062000 &   199100 \nl
\tablevspace{3pt}
0.80 &              1st: & 0.2469 &   0.0764 &   0.0236 &   0.6728 &
                               0.001629 &    100.7 &   999800 &   199300 \nl
\tablevspace{3pt}
0.85 &              1st: & 0.2487 &   0.0752 &   0.0250 &   0.6728 &
                               0.001419 &     51.7 &   810500 &   200000 \nl
\tablevspace{3pt}
0.90 &              1st: & 0.2505 &   0.0722 &   0.0285 &   0.6728 &
                               0.001247 &     43.1 &   478500 &   201800 \nl
\tablevspace{3pt}
1.00 &              1st: & 0.2538 &   0.0663 &   0.0355 &   0.6728 &
                               0.000983 &     39.8 &    85410 &   212600 \nl
     \tablevspace{-3pt}
     &  $\!\!${\bf CBP:} & 0.2538 & 1.78($-3$) & 0.1106 &   0.6729 &
                             2.97($-7$) &     3.52 &    308.7 &   877500 \nl
     \tablevspace{-3pt}
 & 2nd:\tablenotemark{b} & 0.2546 &   0.0563 &   0.0485 &   0.6730 &
                               0.000741 &     27.5 &    239.1 &   170100 \nl
\tablevspace{3pt}
1.10 &              1st: & 0.2566 &   0.0616 &   0.0409 &   0.6729 &
                               0.000799 &     35.5 &    15110 &   227000 \nl
     \tablevspace{-3pt}
     &  $\!\!${\bf CBP:} & 0.2566 & 1.78($-3$) & 0.1106 &   0.6729 &
                             4.16($-7$) &     3.50 &    411.0 &  1204000 \nl
     \tablevspace{-3pt}
 & 2nd:\tablenotemark{b} & 0.2575 &   0.0528 &   0.0588 &   0.6657 &
                               0.000620 &     26.2 &    244.6 &   177900 \nl
\tablevspace{3pt}
1.20 &              1st: & 0.2587 &   0.0578 &   0.0454 &   0.6728 &
                               0.000663 &     33.0 &     3495 &   237000 \nl
     \tablevspace{-3pt}
     &  $\!\!${\bf CBP:} & 0.2587 & 1.95($-3$) & 0.1104 &   0.6729 &
                             6.40($-7$) &     3.37 &    497.4 &  1456000 \nl
     \tablevspace{-3pt}
 & 2nd:\tablenotemark{b} & 0.2597 &   0.0497 &   0.0679 &   0.6606 &
                               0.000526 &     25.2 &    238.9 &   183800 \nl
\tablevspace{3pt}
1.35 &              1st: & 0.2612 &   0.0528 &   0.0512 &   0.6729 &
                               0.000517 &     31.0 &    657.9 &   231300 \nl
     \tablevspace{-3pt}
     &  $\!\!${\bf CBP:} & 0.2612 & 4.26($-3$) & 0.1078 &   0.6729 &
                             1.49($-6$) &     2.81 &    405.1 &   958900 \nl
     \tablevspace{-3pt}
 & 2nd:\tablenotemark{b} & 0.2617 &   0.0467 &   0.0641 &   0.6677 &
                               0.000427 &     24.3 &    229.7 &   192600 \nl
\tablevspace{3pt}
1.50 &              1st: & 0.2626 &   0.0490 &   0.0558 &   0.6730 &
                               0.000418 &     29.5 &    176.5 &   178500 \nl
     \tablevspace{-3pt}
     &  $\!\!${\bf CBP:} & 0.2626 &   0.0120 &   0.0992 &   0.6730 &
                             3.37($-6$) &     2.62 &    162.0 &   331300 \nl
     \tablevspace{-3pt}
 & 2nd:\tablenotemark{b} & 0.2633 &   0.0431 &   0.0755 &   0.6593 &
                               0.000346 &     23.4 &    126.8 &   159600 \nl
\tablevspace{3pt}
1.65 &              1st: & 0.2629 &   0.0460 &   0.0615 &   0.6706 &
                               0.000345 &     27.2 &     87.6 &   130400 \nl
     \tablevspace{-3pt}
     &  $\!\!${\bf CBP:} & 0.2629 &   0.0362 &   0.0736 &   0.6706 &
                             1.71($-5$) &     3.56 &     86.6 &   148900 \nl
     \tablevspace{-3pt}
 & 2nd:\tablenotemark{b} & 0.2635 &   0.0407 &   0.0797 &   0.6577 &
                               0.000289 &     21.8 &     75.4 &   121900 \nl
\tablevspace{3pt}
1.80 &              1st: & 0.2621 &   0.0438 &   0.0708 &   0.6631 &
                               0.000292 &     26.0 &     67.1 &   111300 \nl
     \tablevspace{-3pt}
     &              2nd: & 0.2627 &   0.0387 &   0.0894 &   0.6493 &
                               0.000246 &     20.9 &     60.7 &   106200 \nl
\tablevspace{3pt}
2.00 &              1st: & 0.2603 &   0.0413 &   0.0854 &   0.6497 &
                               0.000241 &     24.4 &     61.0 &   104800 \nl
     \tablevspace{-3pt}
     &              2nd: & 0.2611 &   0.0362 &   0.1062 &   0.6332 &
                               0.000203 &     19.5 &     56.2 &   100900 \nl
\tablevspace{3pt}
2.10 &              1st: & 0.2591 &   0.0403 &   0.0911 &   0.6446 &
                               0.000221 &     23.7 &     59.9 &   103300 \nl
     \tablevspace{-3pt}
     &              2nd: & 0.2599 &   0.0351 &   0.1130 &   0.6268 &
                               0.000187 &     19.0 &     55.5 &    99900 \nl
\tablevspace{3pt}
2.20 &              1st: & 0.2445 &   0.0488 &   0.0559 &   0.6729 &
                               0.000243 &     24.1 &    877.0 &   248600 \nl
     \tablevspace{-3pt}
     &              2nd: & 0.2489 &   0.0431 &   0.0801 &   0.6534 &
                               0.000205 &     19.9 &    261.5 &   203900 \nl
\tablevspace{3pt}
2.25 &              1st: & 0.2396 &   0.0669 &   0.0349 &   0.6728 &
                               0.000289 &     23.7 &   124100 &   207300 \nl
     \tablevspace{-3pt}
     &              2nd: & 0.2483 &   0.0443 &   0.0789 &   0.6532 &
                               0.000217 &     23.0 &    255.5 &   201400 \nl
\tablevspace{3pt}
2.30 &              1st: & 0.2387 &   0.0759 &   0.0242 &   0.6728 &
                               0.000309 &     45.7 &   775300 &   199800 \nl
     \tablevspace{-3pt}
     &              2nd: & 0.2482 &   0.0436 &   0.0799 &   0.6526 &
                               0.000209 &     23.3 &    232.8 &   196300 \nl
\tablevspace{3pt}
     \tablebreak
2.40 &              1st: & 0.2381 &   0.0764 &   0.0234 &   0.6728 &
                               0.000195 &     9417 &  1062000 &   199100 \nl
     \tablevspace{-3pt}
     &              2nd: & 0.2479 &   0.0429 &   0.0818 &   0.6516 &
                               0.000193 &     23.3 &    187.7 &   183300 \nl
\tablevspace{3pt}
2.50 &              1st: & 0.2380 &   0.0765 &   0.0234 &   0.6728 &
                             7.40($-5$) &    35820 &  1062000 &   199100 \nl
     \tablevspace{-3pt}
     &              2nd: & 0.2478 &   0.0420 &   0.0838 &   0.6501 &
                               0.000181 &     21.4 &    155.1 &   171200 \nl
\tablevspace{3pt}
2.75 &              1st: & 0.2380 &   0.0765 &   0.0234 &   0.6728 &
                             7.14($-5$) &    35820 &  1062000 &   199100 \nl
     \tablevspace{-3pt}
     &              2nd: & 0.2478 &   0.0399 &   0.0902 &   0.6455 &
                               0.000156 &     22.0 &    108.1 &   145000 \nl
\tablevspace{3pt}
3.00 &              1st: & 0.2380 &   0.0765 &   0.0234 &   0.6728 &
                             7.14($-5$) &    35820 &  1062000 &   199100 \nl
     \tablevspace{-3pt}
     &              2nd: & 0.2489 &   0.0387 &   0.0965 &   0.6401 &
                               0.000136 &     21.3 &     86.7 &   128500 \nl
\tablevspace{3pt}
3.50 &              1st: & 0.2380 &   0.0765 &   0.0234 &   0.6728 &
                             7.14($-5$) &    35820 &  1062000 &   199100 \nl
     \tablevspace{-3pt}
     &              2nd: & 0.2702 &   0.0363 &   0.1227 &   0.6133 &
                               0.000104 &     19.6 &     67.6 &   110000 \nl
\tablevspace{3pt}
4.00 &              1st: & 0.2380 &   0.0765 &   0.0234 &   0.6728 &
                             7.14($-5$) &    35820 &  1062000 &   199100 \nl
     \tablevspace{-3pt}
     &              2nd: & 0.2883 &   0.0349 &   0.1436 &   0.5914 &
                             8.29($-5$) &     19.5 &     63.4 &   105500 \nl
\tablevspace{3pt}
4.50 &              1st: & 0.2380 &   0.0765 &   0.0234 &   0.6728 &
                             7.14($-5$) &    35820 &  1062000 &   199100 \nl
     \tablevspace{-3pt}
     &              2nd: & 0.3035 &   0.0340 &   0.1628 &   0.5709 &
                             6.88($-5$) &     18.0 &     62.4 &   103400 \nl
\tablevspace{3pt}
5.00 &              1st: & 0.2380 &   0.0765 &   0.0234 &   0.6728 &
                             7.14($-5$) &    35820 &  1062000 &   199100 \nl
     \tablevspace{-3pt}
     &              2nd: & 0.3149 &   0.0336 &   0.1786 &   0.5538 &
                             5.90($-5$) &     18.0 &     63.9 &    81880 \nl
\tablevspace{3pt}
5.50 &              1st: & 0.2380 &   0.0765 &   0.0234 &   0.6728 &
                             7.14($-5$) &    35820 &  1062000 &   199100 \nl
     \tablevspace{-3pt}
     &              2nd: & 0.3247 &   0.0352 &   0.1915 &   0.5397 &
                             5.20($-5$) &     18.8 &     68.8 &    57830 \nl
\tablevspace{3pt}
6.00 &              1st: & 0.2380 &   0.0765 &   0.0234 &   0.6728 &
                             7.14($-5$) &    35820 &  1062000 &   199100 \nl
     \tablevspace{-3pt}
     &              2nd: & 0.3318 &   0.0347 &   0.2007 &   0.5296 &
                             4.65($-5$) &     18.8 &     72.9 &    31500 \nl
\tablevspace{3pt}
6.50 &              1st: & 0.2380 &   0.0765 &   0.0234 &   0.6728 &
                             7.14($-5$) &    35820 &  1062000 &   199100 \nl
     \tablevspace{-3pt}
 & 2nd:\tablenotemark{c} & 0.3368 &   0.0475 &   0.2079 &   0.5217 &
                             4.25($-5$) &     26.0 &     78.0 &     1849 \nl
\tablevspace{3pt}
7.00 &              1st: & 0.2380 &   0.0765 &   0.0234 &   0.6728 &
                             7.14($-5$) &    35820 &  1062000 &   199100 \nl
     \tablevspace{-3pt}
 & 2nd:\tablenotemark{c} & 0.3367 &   0.0396 &   0.2115 &   0.5178 &
                             3.93($-5$) &     21.0 &     82.2 &     4630 \nl
\tablevspace{3pt}
7.50 &              1st: & 0.2380 &   0.0765 &   0.0234 &   0.6728 &
                             7.14($-5$) &    35820 &  1062000 &   199100 \nl
     \tablevspace{-3pt}
 & 2nd:\tablenotemark{c} & 0.2507 &   0.0335 &   0.1550 &   0.5802 &
                             4.16($-5$) &     18.4 &     86.7 &   126900 \nl
\tablevspace{3pt}
8.00 &              1st: & 0.2380 &   0.0765 &   0.0234 &   0.6728 &
                             7.14($-5$) &    35820 &  1062000 &   199100 \nl
     \tablevspace{-3pt}
 & 2nd:\tablenotemark{c} & 0.2474 &   0.0335 &   0.1548 &   0.5805 &
                             3.96($-5$) &     18.3 &     92.2 &   131400 \nl
\tablevspace{3pt}
9.00 &              1st: & 0.2380 &   0.0765 &   0.0234 &   0.6728 &
                             7.14($-5$) &    35820 &  1062000 &   199100 \nl
     \tablevspace{-3pt}
 & 2nd:\tablenotemark{c} & 0.3048 &   0.0320 &   0.1984 &   0.5325 &
                             3.32($-5$) &     15.7 &    101.5 &   135900 \nl
\enddata
     \vskip -15pt

\tablenotetext{a,b,c}{\ As in Table~\protect\ref{tabdrcbplzo}.}

\end{deluxetable}


\end{document}